\documentclass[10pt,journal,compsoc]{IEEEtran}
\ifCLASSOPTIONcompsoc
  \usepackage[nocompress]{cite}
\else
  \usepackage{cite}
\fi
\ifCLASSINFOpdf
\else
\fi
\usepackage{pgfplots}
\pgfplotsset{compat=1.17}

\usepackage{pgfplots} 
\usetikzlibrary{patterns}
\usepackage{booktabs} % For formal tables
\usepackage{url}
\usepackage{xspace}
\usepackage[font=normal]{caption}
\usepackage{subcaption}
\usepackage{float}
\usepackage{textcomp}
\usepackage{multirow,makecell}
\usepackage{pifont}
\usepackage{tikz}
\usepackage{bbm}
\usepackage[inline]{enumitem}
\usepackage{adjustbox}
\newcommand{\etal}{{\em et al.}\xspace}

\newcommand{\BfPara}[1]{{\vspace{1.5mm}\noindent\bf#1.}\xspace}

\usepackage{xcolor}

\usepackage{listings}
\usepackage{tikz}
\usetikzlibrary{tikzmark}
\usetikzmarklibrary{listings}
\usepackage{amsmath}
\usepackage{amssymb}

\usepackage{amsmath}
\usepackage{graphicx}
\usepackage{makecell}

\usepackage{colortbl}
\definecolor{darkgreen}{rgb}{0.0, 0.3, 0.13}
\definecolor{darkred}{rgb}{0.2, 0.0, 0.13}

\usepackage[most]{tcolorbox}
\tcbset{textmarker/.style={%
        parbox=false,boxrule=0mm,boxsep=0mm,arc=0mm,
        outer arc=0mm,left=3mm,right=3mm,top=3pt,bottom=3pt,
        toptitle=1mm,bottomtitle=1mm}
        }%,oversize
        
\newtcolorbox{blueBox}{textmarker,
    colback=blue!10!white}

\newcommand{\takeaway}[1]{\begin{tcolorbox}[
  colback=blue!10!white,
  colbacklower=blue!2!white,
  colframe=blue!70!white,
  title=\textbf{Takeaway},
  fonttitle=\bfseries,
  enhanced,
  size=small,
  sharp corners=south,
  boxrule=0.25pt,  
  top=1pt,
  bottom=1pt,
  left=1pt,
  right=1pt,
  enhanced
]
#1
\end{tcolorbox}}

\usepackage[english]{babel}
\usepackage{blindtext}
\usepackage{xspace}
\usepackage{filecontents}
\usepackage{multicol}

\lstdefinestyle{myStyle}{
  belowcaptionskip=1\baselineskip,
  breaklines=true,
  language=C++,
  showstringspaces=false,
  basicstyle=\footnotesize\ttfamily,
  keywordstyle=\bfseries\color{green!40!black},
  commentstyle=\itshape\color{purple!40!black},
  identifierstyle=\color{blue},
  stringstyle=\color{orange},
  numbers=left,
  firstnumber=1,
}

\usepackage{hyperref}
\usepackage{tikz}
\usepackage{tabularx}
\usepackage{colortbl}
\usepackage{orcidlink}

\hypersetup{
	plainpages=false,
	colorlinks,
	urlcolor=blue,%blue!70!black,
	linkcolor=blue,%red!70!black,
	citecolor=blue,%green!70!black,
	bookmarksnumbered
}

\tcbset{
  condensedtakeaway/.style={
    colback=blue!5,
    colframe=blue!75!black,
    fonttitle=\bfseries,
    title=Takeaway,
    boxrule=0.8pt,
    arc=2mm,
    left=2mm,
    right=2mm,
    top=1mm,
    bottom=1mm,
    enhanced,
    sharp corners=south,
  }
}

\begin{document}

\title{Understanding Concept Drift with Deprecated Permissions in Android Malware Detection}

\author{Ahmed Sabbah\orcidlink{0000-0001-5034-8038}, Radi Jarrar\orcidlink{0000-0003-2692-8096}, Samer Zein\orcidlink{0000-0003-3720-4384},  David~Mohaisen\orcidlink{0000-0003-3227-2505}\IEEEmembership{, Senior Member,~IEEE}

\IEEEcompsocitemizethanks{\IEEEcompsocthanksitem Ahmed Sabbah, Radi Jarrar, and Samer Zein are with the Department of Computer Science, University of Birzeit, Palestine.
E-mail: asabah@birzeit.edu, rjarrar@birzeit.edu, szain@birzeit.edu. \IEEEcompsocthanksitem David Mohaisen is with the Department of Computer Science, University of Central Florida, Orlando, FL 32816 USA. E-mail: mohaisen@ucf.edu (Corresponding author).}} %\thanks{Manuscript received April 19, 2005; revised August 26, 2015.}

%\markboth{IEEE Transactions on Dependable and Secure Computing,~Vol.~14, No.~8, August~2015}%{Sabbah \MakeLowercase{\textit{et al.}}: Understanding Concept Drift with Deprecated Permissions in Android Malware Detection}

\IEEEtitleabstractindextext{%
\begin{abstract}
Permission analysis is a widely used method for Android malware detection. It involves examining the permissions requested by an application to access sensitive data or perform potentially malicious actions. In recent years, various machine learning (ML) algorithms have been applied to Android malware detection using permission-based features and feature selection techniques, often achieving high accuracy. However, these studies have largely overlooked important factors such as protection levels and the deprecation or restriction of permissions due to updates in the Android OS—factors that can contribute to concept drift.

In this study, we investigate the impact of deprecated and restricted permissions on the performance of machine learning models. A large dataset containing 166 permissions was used, encompassing more than 70,000 malware and benign applications. Various machine learning and deep learning algorithms were employed as classifiers, along with different concept drift detection strategies. The results suggest that Android permissions are highly effective features for malware detection, with the exclusion of deprecated and restricted permissions having only a marginal impact on model performance. In some cases, such as with CNN, accuracy improved. Excluding these permissions also enhanced the detection of concept drift using a year-to-year analysis strategy. Dataset balancing further improved model performance, reduced low-accuracy instances, and enhanced concept drift detection via the Kolmogorov–Smirnov test.
\end{abstract}

}

\maketitle

\IEEEdisplaynontitleabstractindextext

\IEEEpeerreviewmaketitle

\section{Introduction}\label{sec:introducion}
Mobile devices are an essential tool in everyday life, providing users with access to a wide range of applications for communication, banking, entertainment, and productivity. Two operating systems dominate the mobile market, Google Android and Apple iOS, with Android taking 71\% of the market share by 2024~\cite{SmartphoneShare:online}. Android employs a permission-based security model that grants applications specific privileges to regulate access to sensitive resources. These permissions control access to hardware components (e.g., camera, microphone, etc.), user data (e.g., contacts, messages, location), and system functionalities (e.g., network access, storage, background processes)~\cite{Permissi97:online}. Android permissions aim to balance security and usability, ensuring that applications can only access resources explicitly allowed by the user.

While the permission system is designed to protect user privacy and device security, permissions can be misused, intentionally by adversaries~\cite{SharmaA24} or unintentionally by novice developers who may request permissions without fully understanding their implications~\cite{CarliniFW12}. This is clear in the finding of more 26\% of more than 83k applications examined that contained at least one bad practice of custom permission issue~\cite{ZhangYLZSZD}. However, while it is important to investigate the bad practices of using permissions, this study focuses on intentional misuse by attackers by requesting excessive permissions to access user data or control device functionality. For example, some malware applications masquerade themselves as legitimate apps, such as flashlights or weather apps, but request permission to read SMS messages, access call logs, or track user locations.

Due to the risks associated with permissions and their role in granting access to user and system resources, permissions are a target for adversaries and third-party apps, enabling them to control devices and access sensitive data. This, in turn, allows defenders to extract permission-based features for malware detection with machine learning algorithms~\cite{IlhamAA18, KatoSS21, MatAKAF2022, SahinKAK23}. Many studies have recognized the importance of permissions and have focused on improving the performance of detection models that use them~\cite{IlhamAA18, ShatnawiYY22}, often through feature selection and optimization~\cite{SahinKAK23, MatAKAF2022, KatoSS21}. Other research has explored combining permissions with additional features, such as APIs, to enhance performance~\cite{MillarMRM21, FangHLLC0HZ23}, yielding strong results. However, most of these studies assume that permissions and their usage patterns remain static over time, failing to consider how changes in permissions impact detection.

Concept drift is closely associated with machine learning models, where performance degrades over time due to changes in the underlying data distribution. Researchers have investigated concept drift from various perspectives. For instance, Hoens~\etal reviewed methods for detecting concept drift and highlighted the challenge of class imbalance in data streams~\cite{HoensPC12}. Ditzler~\etal examined existing methods for detecting concept drift, considering both active and passive approaches~\cite{DitzlerRAP15}. Webb~\etal introduced formal definitions of concept drift and proposed a taxonomy that includes various types, such as drift duration, drift transition, and drift recurrence~\cite{WebbHCNP16}. Krawczyk~\etal surveyed ensemble learning methods for data stream classification and regression tasks in the context of concept drift~\cite{KrawczykMGSW17}. Moreover, some studies have focused on adaptation methods in machine learning to address concept drift~\cite{BayramAK22}.

In the context of Android malware detection, concept drift has been gaining momentum, due to ML widespread adoption in the defense-attack arms race. Ensemble classifiers with sliding windows and feature selection are used to adaptively detect malware in streaming data~\cite{HuMZLYL17}. Retraining methods used for concept drift detection and sampling methods to maintain detector performance in changing environments~\cite{MolinaCoronadoMMM23}. Self-training with pseudo-labels is adapted to handle shifting data distributions, mitigating concept drift, and reducing annotation efforts when combined with active learning~\cite{AlamFMR24}. Chow~\etal proposed a framework for analyzing dataset drift exploring root causes~\cite{ChowKLCAP23}. 

Despite these findings, concept drift studies are often limited to the natural evolution of usage patterns while ignoring an important aspect of evolution: {\bf deprecation}. Since permissions are continuously added, deprecated, or restricted, models trained on outdated permissions may become ineffective, leading to a decline in detection accuracy. Malware detection models trained on older data may struggle to accurately classify newer malware due to changes in permission usage patterns. This issue has been largely neglected in existing research, despite its critical implications for the long-term effectiveness of malware detection.

In this work, we investigate the impact of Android permission evolution on malware detection by analyzing how deprecated and restricted permissions contribute to concept drift. We evaluate the effectiveness of permissions as features in malware classification by training models such as Random Forest (RF), Convolutional Neural Networks (CNN), and Recurrent Neural Networks (RNN) on historical data and testing them on newer datasets. Additionally, we assess how removing deprecated and restricted permissions affects model performance over time. By quantifying concept drift in both traditional machine learning and deep learning models, we provide insights into the challenges of maintaining effective malware detection systems in the context of a rapidly evolving Android permission landscape.

\BfPara{Contributions} We make the following contributions:  
\begin{enumerate}[leftmargin=*]
\item {\em Analyzing the use of permissions in malware and benign applications.}  We examine how permissions are used in malicious and benign applications, identifying patterns and differences that can inform effective permission abuse prevention and malware detection strategies 

\item {\em Investigating the impact of permission evolution on concept drift.}  We analyze the effect of permission changes over time by training models on historical datasets and testing them on newer data. For example, we train on permissions from 2008 and evaluate on data from subsequent years (2009–2020).  

\item {\em Examining the effect of deprecated and restricted permissions on model stability.}  
We assess how the removal of deprecated and restricted permissions influences malware detection performance, highlighting potential pitfalls in using outdated or evolving feature sets.    
\end{enumerate}

\BfPara{Organization} The background is outlined in section~\ref{sec:Background}, followed by the methodology in section~\ref{sec:Methodology}, the results and discussion in section~\ref{sec:Experiments}, the related work in section~\ref{sec:related}, followed by concluding remarks in section~\ref{sec:conclusion}.

\section{Background}
\label{sec:Background}
Android malware apps can be installed via official stores, third parties, or social engineering tactics \cite{MengPXLZ19} to gain unauthorized access and exploit root privileges without user consent \cite{FeltFCHW11}. Android malware varies widely and can be categorized based on shared characteristics or behaviors. Common types include banking malware, Trojans, spyware, and worms, with each type further classified into families based on specific traits or attack patterns.

\subsection{Android Package Kit (APK)}
The Android Package Kit (APK) is an archive file format used to deploy and install applications on the Android operating system. APK files can be downloaded and installed from the official Google Play Store or from third-party sources. To install APKs from outside the official app store, Android users must enable a setting feature that allows installations from unknown sources. The main components of the APK file are as follows:
\begin{itemize}[leftmargin=*] 
    \item {\tt AndroidManifest.xml}: This file contains XML metadata information about components of the application, such as security permissions, activities, and services.
    
    \item {\tt Classes.dex}: This file contains the source code of an application written in Java and compiled to a Dalvik executable with \textit{.edx} extension.
    
    \item {\tt Resources.arsc}: A binary XML file that includes precompiled application resources~\cite{Ali19}.
    
    \item {\tt Resources (res/)}: A folder containing non-compiled resources that the application requires at run time, such as menus, images, layouts, and database use.
    
    \item {\tt Assets (assets/)}: Optional folder containing application assets that Asset Manager can retrieve.
    
    \item {\tt Libraries (lib/)}: Optional folder containing code compiled for various processors; e.g., ARM and x86~\cite{Ali19}.
    
    \item {\tt META-INF}: Folder containing the {\tt MANIFEST.MF} file. Further, it includes the application developer's signature, which can be used to authenticate an external developer.
\end{itemize}

\subsection{Android Security}
As an open-source platform, Android has been an ongoing target of hackers. Consequently, ensuring the security of applications remains a critical concern and a significant challenge. To mitigate this risk, Android OS implements a two-level security framework: one at the Linux kernel level and another at the application level~\cite{VishnoiMNP21}. The security mechanisms of Android include the following features: 
\begin{itemize}[leftmargin=*]
    \item \textbf{Sandbox:} Each application is assigned a unique user ID (UID) and runs in its own process and isolated space. Thus, one application with its own UID did not allow it to be accessed or modified by another application. Additionally, the application running in its own sandbox has limited access to system resources. 
    
    \item \textbf{Permissions:} Any application must define the permission needs of other application components or Android resources in the AndroidManifest.xml file. The user must grant these permissions during app installation; otherwise, the application cannot be installed on the device. Moreover, permission can be granted during the application runtime, which is called dynamic permissions.
    
     \item \textbf{Signatures:} The Android application must be stamped and digitally signed with a certificate that identifies the developer of the application. A developer with the same certificate can update the application in the future. In addition, applications with this signature can trust each other by sharing the UID between them.
     
     \item \textbf{Secure Inter-process Communication (IPC):} IPC protocols allow an application to communicate with remote servers and also other applications.
\end{itemize}

\subsection{Android Permissions}
\begin{figure}[t]
    \centering
    \includegraphics[scale=0.55]{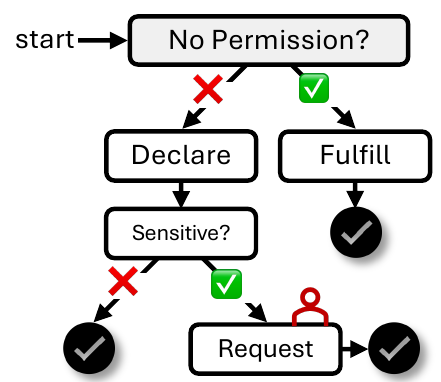}
    \caption{Workflow for Android permissions.}
    \label{fig:PermOnline}\vspace{-3mm}
\end{figure}

Android apps are required to request explicit permission from users before accessing sensitive resources or data, such as the camera, location, or microphone. This ensures that applications cannot access personal information without the user's knowledge and explicit consent. The workflow of using app permissions is illustrated in \autoref{fig:PermOnline}.

The Android operating system provides normal, dangerous, signature, custom, install-time, and special permissions~\cite{Permissi97:online}, which we highlight in the following: 

\begin{enumerate} [leftmargin=*] 
\item \textbf{Normal Permissions:} These permissions are automatically granted during installation and are associated with basic functionalities like internet access, Bluetooth, and NFC. They pose minimal privacy and security risks as they do not access sensitive data.

\item \textbf{Dangerous Permissions:} Require explicit user approval during execution as they grant access to sensitive data or critical functions, such as contacts, location, microphone, and camera, posing potential privacy and security risks.  

\item \textbf{Signature Permissions:} These permissions enable apps signed with the same certificate (private key) to share privileged information or resources. New apps are automatically granted these permissions if signed with the same certificate as existing applications.  

\item \textbf{Custom Permissions:} These permissions are defined by developers to enforce access control between apps. Moreover, these permissions can be classified as normal or dangerous depending on the level of access they provide. This type is widely used but often undocumented and hidden from users, posing security risks by enabling indirect access to protected resources without consent~\cite{GambaFBBRTV24}.

\item \textbf{Install-Time Permissions:} Introduced in Android 6.0, these permissions are granted by users during installation instead of being requested at runtime.  

\item \textbf{Special Permissions:} These permissions are restricted to system apps or those signed with the same certificate as system applications. Moreover, these permissions are not available to third-party apps and must be explicitly granted by the platform or device manufacturer.  
\end{enumerate}

\subsection{Permission-Based Malware Detection}
Android malware detection using permission-based analysis involves multiple stages to extract, preprocess, and classify apps based on their permissions. The pipeline can be structured into the following key components:

\BfPara{Data Collection and Preprocessing} The android malware detection begins with collecting APK files from various sources, including official stores, third-party repositories, and malware datasets. Each APK undergoes static analysis to extract requested permissions from the \texttt{AndroidManifest.xml} file. Let \( A = \{a_1, a_2, \dots, a_n\} \) represent a set of applications, and each app \( a_i \) requests a subset of permissions \( P_i = \{p_1, p_2, \dots, p_m\} \), where \( P \) is the global set of all possible permissions.

\BfPara{Feature Engineering and Selection} Extracted permissions serve as features for classification. Given an application \( a_i \), its permission vector \( v_i \) can be represented as $v_i = (x_1, x_2, \dots, x_m),$ where \( x_j \in \{0,1\} \) indicates whether permission \( p_j \) is requested or not in the application.

\BfPara{Training and Classification} A machine learning or deep learning classifier is trained using labeled data. Given a training dataset \( D = \{(v_1, y_1), (v_2, y_2), \dots, (v_n, y_n)\} \), where \( y_i \in \{0,1\} \) denotes whether an app is benign (\( 0 \)) or malicious (\( 1 \)), the model learns a mapping function $f: v \rightarrow y$. 

Classification algorithms used in permission-based malware detection include {\em traditional machine learning algorithms}, such as random forest (RF), support vector machines (SVM), decision trees (DT), and {\em deep learning algorithms}, such as neural networks (NN), long short-term memory (LSTM), and graph neural networks (GNN).

\subsection{Concept Drift}
The phenomenon of concept drift refers to the unexpected change in the statistical properties or defining features of the target variable over time in non-stationary data distributions~\cite{SchlimmerG86}. Mathematically, let \( P_t(X, Y) \) represent the joint probability distribution of the input features \( X \) and the target variable \( Y \) at time \( t \). Concept drift occurs when: $P_{t_1}(X, Y) \neq P_{t_2}(X, Y), \quad \text{for } t_1 \neq t_2.$ This change can manifest in different forms: abrupt, incremental, gradual, and recurring~\cite{XiangZCW23}. Each form corresponds to distinct transitions in the data distribution:

\BfPara{Abrupt Drift} A sudden shift in the data distribution at a specific point in time, mathematically defined as:
$P_{t}(X, Y) =
\begin{cases}
P_A(X, Y), & t < t_c \\
P_B(X, Y), & t \geq t_c
\end{cases}$, where \( t_c \) is the time at which the drift occurs, and \( P_A \) and \( P_B \) represent different distributions before and after the drift.

\BfPara{Incremental Drift} In this type of drift, the data distribution evolves progressively over time, making the transition smooth, expressed as follows:
\begin{eqnarray}\nonumber
\nonumber P_{t}(X, Y) = \alpha_t P_A(X, Y) + (1 - \alpha_t) P_B(X, Y), \\
\nonumber \quad 0 < \alpha_t < 1, \quad \lim_{t \to \infty} \alpha_t = 0.
\end{eqnarray}

\BfPara{Gradual Drift} In this type, both old and new distributions coexist during a transition period before the new distribution fully replaces the old one:
\begin{equation}\nonumber
P_{t}(X, Y) =
\begin{cases}
(1 - \beta_t) P_A(X, Y) + \beta_t P_B(X, Y), & t_c \leq t \leq t_d \\
P_B(X, Y), & t > t_d
\end{cases},
\end{equation}
where \( \beta_t \) gradually increases from 0 to 1 over the transition period \( [t_c, t_d] \) leading to the gradual drift.

\BfPara{Recurring Drift} In this type, a previously observed concept reappears at later time steps:
\begin{eqnarray}\nonumber
\nonumber P_{t}(X, Y) \in \{P_A(X, Y), P_B(X, Y), P_C(X, Y), \dots \}, \quad \\
\nonumber \text{where } P_t(X, Y) \text{ cycles over time}.
\end{eqnarray}

In conclusion, abrupt drift involves rapid concept transitions, incremental drift refers to slower changes, gradual drift involves periodic concept shifts, and recurring drift is characterized by the reappearance of earlier concepts~\cite{XiangZCW23}. These forms of concept drift indicate changes in the underlying data distribution, necessitating adaptive learning models to maintain performance in dynamic environments.

\begin{figure*}[t]
    \centering
    \includegraphics[width=0.99\linewidth]{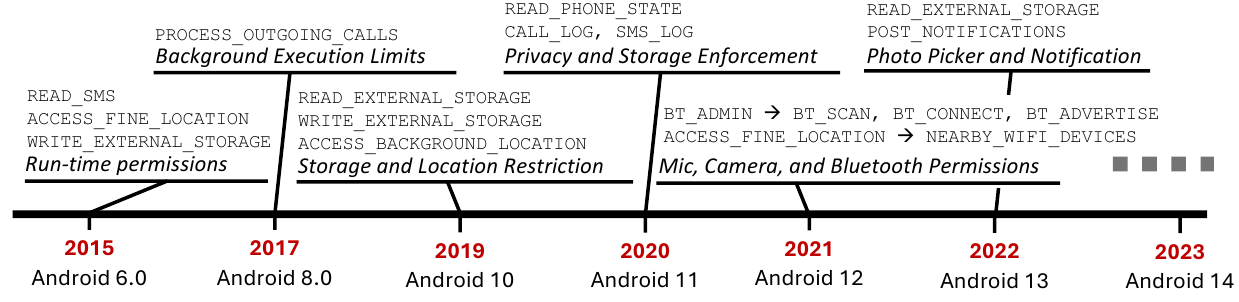}\vspace{-3mm}
    \caption{A timeline of the Android permission system evolution and deprecation across various versions. The permissions are added or removed, and Android 14 and 15 (2023 and 2024) were removed due to the lack of space.}
    \label{fig:dep}\vspace{-3mm}
\end{figure*}

\subsection{Android Permissions Deprecation}

The Android permission system is not static and continuously evolves as new hardware is introduced, security policies change, and user expectations shift. Consequently, Google has introduced new permissions, deprecated old ones, and restricted access to certain sensitive permissions to enhance security~\cite{Permissi97:online}, as shown in the timeline in \autoref{fig:dep}. Deprecated permissions are those removed or replaced with more secure alternatives. For example, the {\tt GET\_TASKS} permission, which allowed applications to retrieve information about running tasks, was deprecated in Android 5.0 (API Level 21) due to privacy risks~\cite{Activity74:online}. Similarly, {\tt READ\_CALL\_LOG} was restricted in Android 9.0 (API Level 28) to prevent unauthorized access to call history~\cite{Activity74:online}. Restricted permissions are those moved to higher security levels, meaning only system or explicitly approved apps can use them. For example, Android 10 (API Level 29) introduced restrictions on {\tt ACCESS\_BACKGROUND\_LOCATION}, limiting its use to prevent apps from stealthily tracking users. Additionally, changes were made to prevent silent access to the device's screen contents by restricting the scope of {\tt READ\_FRAME\_BUFFER} to signature access~\cite{Privacyc75:online}.

\section{Methodology}\label{sec:Methodology}
Concept drift occurs when the relationship between features and target labels changes over time~\cite{GamaZBPB14}. In the context of Android malware detection, the continuous evolution of the Android permission system--driven by security updates, hardware advancements, and changes in app development practices--affects the reliability of permission-based detection models. Although permissions have been widely used as features in machine learning models for malware detection, prior studies have largely treated them as static attributes, overlooking how their deprecation, restriction, or modification over time can impact model effectiveness.

We aim to bridge this gap by: (1) reevaluating the effectiveness of permissions as features for machine learning and deep learning models, (2) assessing their sensitivity to concept drift as permissions evolve, and (3) analyzing the impact of deprecated and restricted permissions on model performance. Consequently, we seek to answer the following research questions.

\begin{itemize}[leftmargin=*]
\item {\bf RQ-1:} How do permission-based features impact malware detection accuracy across ML and DL models?  

\item {\bf RQ-2:} How does permission sensitivity to concept drift affect Android malware detection over time?  

\item {\bf RQ-3:} How do deprecated and restricted permissions influence model stability, accuracy, and drift?

\end{itemize}

To achieve the study goals and answer the questions, we propose the following strategies as steps of exploration:

\begin{enumerate}[leftmargin=*]
\item \textbf{Ignoring the Chronological Context.} Algorithms were tested on all features, including timestamps.  

\item \textbf{Year-to-Year Strategy.} Models were trained on data from one year and tested on subsequent years (e.g., trained on 2008, tested on 2009–2020).  

\item \textbf{Exclusion Strategy.} Specific permission categories—deprecated (D), restricted (R), and not-for-use-by-third-party (N)—were systematically removed to assess their impact on malware detection and concept drift. Results were compared against a baseline using all permissions (ALL) in both prior strategies.
\end{enumerate}

\subsection{Model Workflow}
In this work, this test was applied to a simulated dataset for concept drift detection \cite{ZhixiongWQWJXZB20}. The dataset was divided into 12 blocks, sequentially labeled, and the $p$-values were calculated over time. We used this test to compute the Cumulative CDFs of two sample groups (e.g., CDFs for the accuracy of training years 2008 and 2010). The KS test measures the maximum absolute difference between the CDFs, and pairwise comparisons were performed by computing the KS test statistic and the corresponding $p$-value for each pair of training years (T1, T2). A significant $p$-value $(<0.05)$ indicates a possible drift between distributions.

\begin{figure}[t]
    \centering
    \includegraphics[scale=0.45]{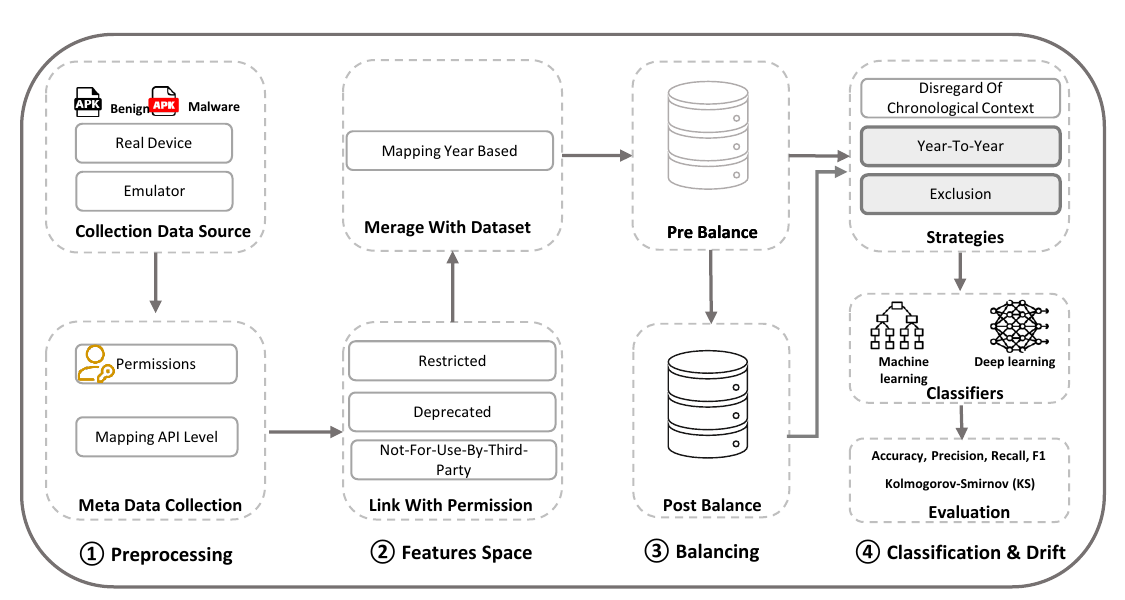}
    \caption{A four-stage framework for Android malware detection and drift analysis: data preprocessing, API-level permission-based feature extraction, dataset balancing, and classification using machine and deep learning. Concept drift is identified using KS statistics and standard metrics.}
    \label{fig:PermPip}\vspace{-3mm}
\end{figure}

\autoref{fig:PermPip} presents a four-stage workflow of this work. In the \textbf{pre-processing} phase, benign and malware are collected from real devices and emulators, followed by the extraction of permissions and their corresponding API levels to collect metadata. In the \textbf{feature space} stage, permissions are mapped to categories such as restricted, deprecated, or not-for-use-by-third-party, and merged with the dataset based on API release year. The \textbf{balance phase} applies the balance algorithm. Finally, in the Classification and Drift phase, temporal evaluation strategies, year-to-year, and exclusion-based are applied using machine learning and deep learning classifiers, with performance evaluated through metrics like accuracy, precision, recall, F1 score, and drift detection via the Kolmogorov–Smirnov (KS) test.

\subsection{Dataset Overview}\label{sec:Dataset Overview}
\begin{figure*}[t]
    \centering
          \begin{tikzpicture}
        \begin{axis}[
            ybar,
            width=18cm,
            height=3cm,
            %ymode=log, % Log-scale for y-axis
            %xlabel={Year},
            ylabel={\# Apps},
            symbolic x coords={2008, 2009, 2010, 2011, 2012, 2013, 2014, 2015, 2016, 2017, 2018, 2019, 2020},
            xtick=data,
            x tick label style={rotate=45, anchor=east},
            ymin=0,
            bar width=5pt,
            ymajorgrids=true,
            enlargelimits=false,
            tick label style={font=\scriptsize},
            legend style={at={(.75,1.4)}, anchor=north, legend columns=2,font=\small},
            nodes near coords,  % Adds numbers on top of bars
            every node near coord/.append style={font=\tiny, rotate=90, anchor=west}, % Rotates numbers vertically
            enlarge x limits=0.05
        ]
        
        % Malware=0 Data
        \addplot[ybar, color=green!70!black, pattern=north east lines, pattern color=green!70!black] coordinates {
            (2008, 64) (2009, 622) (2010, 5074) (2011, 22873) 
            (2012, 342) (2013, 489) (2014, 632) (2015, 720) 
            (2016, 743) (2017, 650) (2018, 775) (2019, 200) (2020, 275)
        };
        \addlegendentry{Benign--Real Device}
        
        % Malware=1 Data
        \addplot[ybar, color=black, pattern=north west lines, pattern color=black] coordinates {
            (2008, 934) (2009, 20) (2010, 269) (2011, 3137) 
            (2012, 7564) (2013, 7487) (2014, 8005) (2015, 1424) 
            (2016, 2445) (2017, 4806) (2018, 4006) (2019, 1491) (2020, 1048)
        };
        \addlegendentry{Malware--Real Device}

                % Malware=0 Data
        \addplot[ybar, color=red, pattern=grid, pattern color=red] coordinates {
            (2008, 66) (2009, 607) (2010, 4978) (2011, 22146) 
            (2012, 309) (2013, 444) (2014, 596) (2015, 689) 
            (2016, 715) (2017, 788) (2018, 733) (2019, 180) (2020, 1002)
        };
        \addlegendentry{Benign--Emulator}
        
        % Malware=1 Data
        \addplot[ybar, color=blue!60,pattern=crosshatch, pattern color=blue] coordinates {
            (2008, 686) (2009, 19) (2010, 260) (2011, 2277) 
            (2012, 6498) (2013, 5641) (2014, 6286) (2015, 1031) 
            (2016, 1941) (2017, 625) (2018, 2299) (2019, 1388) (2020, 256)
        };
        \addlegendentry{Malware--Emulator}
        \end{axis}
    \end{tikzpicture}
        \vspace{-3mm}

    \caption{KronoDroid dataset distribution of both malware and benign samples across different years.}
    \label{fig:Kdataset}\vspace{-3mm}
\end{figure*}
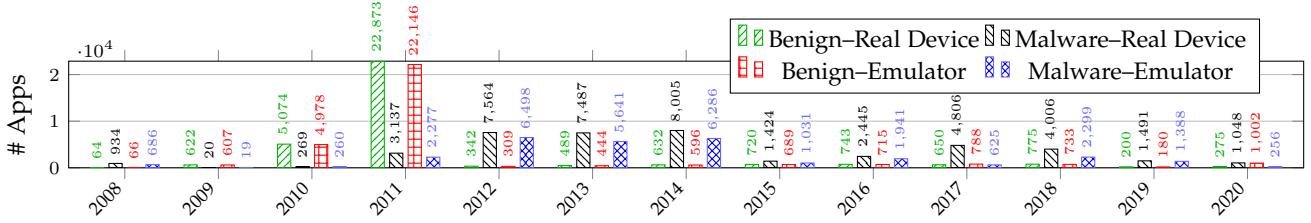

The Kronodroid dataset \cite{Guerra-ManzanaresBN21} was built using a combination of static and dynamic features extracted from Android applications. The source of the malware samples was selected from different databases: VirusTotal, Drebin, VirusShare, and AMD. The benign apps were selected from APKMirror, F-droid, and MARVIN. This dataset includes 489 dynamic and static features extracted from Android applications that covered the years 2008 to 2020. The emulator dataset contains 28,745 malware samples and 35,246 benign samples. The real device dataset includes 41,382 malware samples and 36,755 benign samples. \autoref{fig:Kdataset} shows the number of malware and benign applications for each year. For permissions features, 166 features were selected from the real device and emulator datasets. An app takes a value of 1 when requesting permission and 0 otherwise. Kronodroid did not provide different levels of protection for each permission, such as normal, signature, and dangerous.  

\subsection{Data Analysis}
\begin{figure}[t]
    \centering
    \includegraphics[scale=0.40]{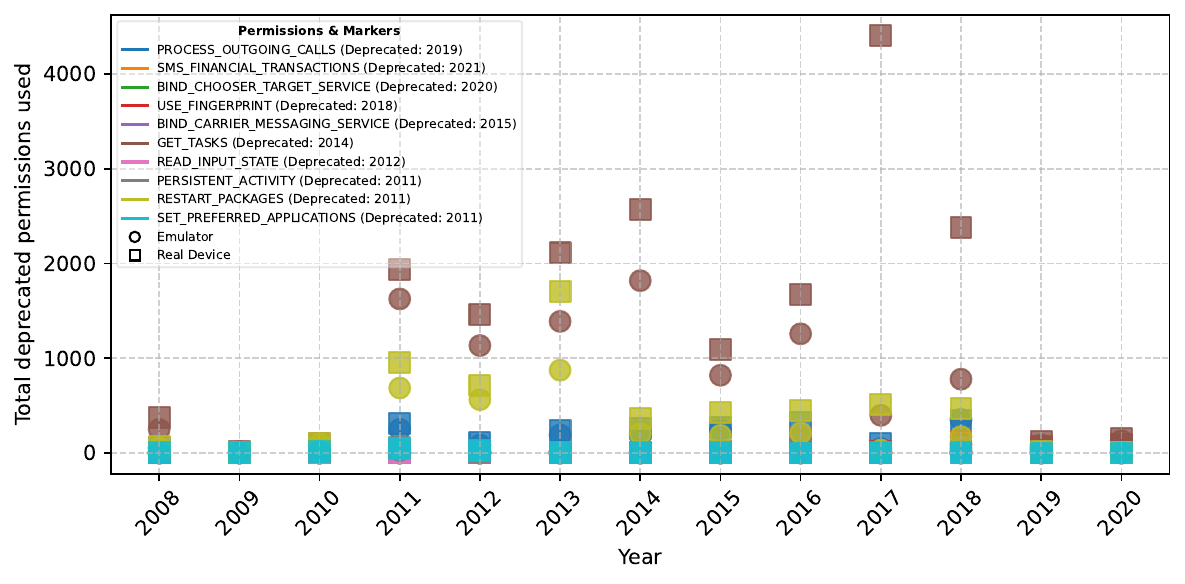}\vspace{-3mm}
    \caption{Deprecated permissions over time--the legend with the deprecation years for real and emulator datasets.}
    \label{fig:DeprecatedEmulatorReal}\vspace{-3mm}
\end{figure}

Because Knorodoid lacks information about the permissions themselves, we addressed this gap by collecting metadata for each permission from the official Android website~\cite{Manifest2:online}. This metadata includes critical attributes, such as the four types of protection levels (normal, dangerous, signature, and not for use by third parties). Additionally, it specifies the API levels at which permissions were deprecated or restricted, allowing them to be mapped to the corresponding years. The usage of deprecated permissions over time for both the emulator and real device is shown in~\autoref{fig:DeprecatedEmulatorReal}.

\BfPara{Obsolete Permissions Induce Drift} Although some permissions were deprecated in earlier years, such as {\tt GET\_TASK}, which was deprecated in 2014, they continued to be used in subsequent years. However, the usage of this permission decreased significantly, particularly in 2019 and 2020. This decline suggests that the associated features became obsolete, contributing to changes in the feature space distribution and potentially causing drift.

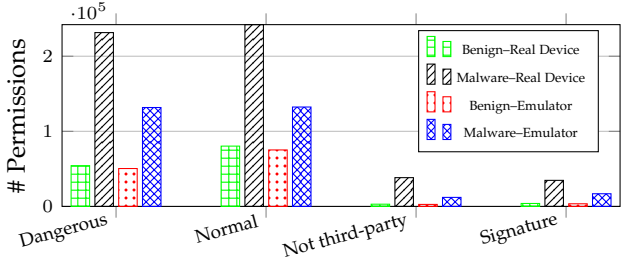
\begin{figure}[t]
    \centering
    \begin{tikzpicture}
        \begin{axis}[
            ybar,
            width=9cm,
            height=4cm,
            bar width=7pt,
            symbolic x coords={Dangerous, Normal, Not third-party, Signature},
            xtick=data,
            x tick label style={rotate=15, anchor=east, font=\scriptsize}, 
            ymin=0,
            ylabel={\# Permissions},
            ymajorgrids=true,
            enlargelimits=false,
            tick label style={font=\scriptsize},
            legend style={at={(.80,0.97)}, anchor=north, legend columns=1,font=\tiny},
            % legend cell align={left}
              enlarge x limits=0.12
        ]
    \addplot[ybar, fill=blue!50, draw=green, pattern=grid, pattern color=green] coordinates {(Dangerous,54280) (Normal,80302) (Not third-party,2967) (Signature,3933)};
\addlegendentry{Benign--Real Device}
\addplot[ybar, fill=red!50, draw=black, pattern=north east lines, pattern color=black] coordinates {(Dangerous,231465) (Normal,242064) (Not third-party,38309) (Signature,34709)};

\addlegendentry{Malware--Real Device}
\addplot[ybar, fill=gray!50, draw=red, pattern=dots, pattern color=red] coordinates {(Dangerous,50448) (Normal,75210) (Not third-party,2670) (Signature,3514)};

\addlegendentry{Benign--Emulator}
\addplot[ybar, fill=blue!70, draw=blue, pattern=crosshatch, pattern color=blue] coordinates {(Dangerous,131710) (Normal,132357) (Not third-party,12018) (Signature,16880)};

\addlegendentry{Malware--Emulator}

        \end{axis}
    \end{tikzpicture}\vspace{-4mm}
    \caption{Protection level usage: malware vs. benign.}
    \label{fig:histogram_real_vs_emu}\vspace{-5mm}
\end{figure}

\BfPara{Malware Exploits Restricted Permissions} To investigate the usage of protection level types by malware, we categorized benign and malicious applications based on their use of permissions. Malware tends to use permissions more extensively than benign applications. However, certain permissions--such as those classified as {\it ``not for use by third parties''}--show notable usage by malware in the emulator dataset. These permissions are generally restricted to pre-installed apps and require the requesting app to share the same digital signature as the app granting the permission. The ability of malware to access these permissions indicates potential exploitation of pre-installed apps and warrants further investigation. \autoref{fig:histogram_real_vs_emu} shows that malware frequently exploits dangerous permissions, highlighting the importance of monitoring these permissions for security.

After analyzing the dataset, we can summarize the number of permissions present and how their usage has changed over time. \autoref{tab:permissions_summary} summarizes the classification of permissions based on their metadata, including their protection levels, number of deprecations, and restriction levels for 166 permissions in the dataset.

\begin{table}[t]
\centering
\caption{Summary of Permission Categories. (N) denotes ``not for use by third-party."}
\label{tab:permissions_summary}\vspace{-2mm}
\begin{tabular}{lccc}
\toprule
\textbf{Category}       & \textbf{Permissions} & \textbf{Deprecated} & \textbf{Restricted} \\ 
\midrule
Dangerous               & 30                   & 1                  & 9                   \\
Normal                  & 54                   & 6                  & -                   \\
Signature               & 43                   & 2                  & -                   \\
(N). Third-Party & 39                   & 1                  & -                   \\
\midrule
\textbf{Total}          & \textbf{166}         & \textbf{10}        & \textbf{9}          \\
\bottomrule
\end{tabular}\vspace{-3mm}
\end{table}

\begin{table}[ht]
\centering
\caption{Protection levels of permissions across Android API levels and years. {\bf Dang}erous, {\bf Int}ernal, {\bf Norm}al, {\bf Sig}nature, Not for use by 3rd party ($\neg$TP), and total are the metrics.}
\label{fig:permissions_New_to2024}\vspace{-2mm}
%\scriptsize
\begin{tabular}{ccccccc}
\hline
\textbf{Year (API)} & \textbf{Dang.} & \textbf{Int} & \textbf{Norm} & \textbf{$\neg$TP} & \textbf{Sig} & \textbf{Total} \\
\hline
2008 (1) & 17 & 0 & 26 & 25 & 6 & 74 \\
2009 (2) & 2 & 0 & 0 & 2 & 0 & 4 \\
2010 (8) & 1 & 0 & 1 & 1 & 2 & 5 \\
2011 (11) & 0 & 0 & 0 & 0 & 1 & 1 \\
2012 (16) & 3 & 0 & 0 & 0 & 1 & 4 \\
2013 (18) & 2 & 0 & 0 & 2 & 1 & 5 \\
2014 (20) & 1 & 0 & 0 & 0 & 0 & 1 \\
2015 (22) & 0 & 0 & 1 & 0 & 0 & 1 \\
2016 (24) & 0 & 0 & 1 & 0 & 3 & 4 \\
2017 (26) & 2 & 0 & 4 & 0 & 3 & 9 \\
2018 (28) & 1 & 0 & 3 & 0 & 0 & 4 \\
2019 (29) & 3 & 0 & 3 & 0 & 4 & 10 \\
2020 (30) & 0 & 0 & 5 & 0 & 3 & 8 \\
2021 (31) & 4 & 0 & 8 & 1 & 4 & 17 \\
2022 (32) & 0 & 0 & 1 & 0 & 0 & 1 \\
2023 (34) & 1 & 79 & 21 & 1 & 2 & 104 \\
2024 (35) & 0 & 6 & 4 & 2 & 0 & 12 \\
Unknown (0) & 0 & 0 & 1 & 0 & 0 & 1 \\
\hline
\end{tabular}
\end{table}

\BfPara{Evolving Permissions Challenge Models} Per the official Android website, the total number of permissions reached 323 at the time of writing this paper. However, our dataset covers the period from 2008 to 2020, with only 166 permissions utilized up to API level 30. The new permissions introduced after 2021, as illustrated in \autoref{fig:permissions_New_to2024}, highlight the evolving nature of Android's security policies. These emerging permissions introduce previously unseen features that were absent in earlier years, posing a significant challenge for machine learning models trained on historical data. Since the model has not encountered these permissions before, its decision boundaries may become less reliable, leading to performance degradation and concept drift.

\subsection{Evaluation Metrics}

\BfPara{Models Performance} We utilize standard classification metrics to evaluate various aspects of model performance. These metrics are used to evaluate the effectiveness of the detection model and monitor performance changes when the model encounters concept drift. The key evaluation metrics are as follows. 
(1)   \textbf{Precision:} Measures the accuracy of correctly classified malware apps:
    $\text{Precision} = \frac{\sum_{i=1}^{K} \text{TP}_i}{\sum_{i=1}^{K} (\text{TP}_i + \text{FP}_i)}$,  
    where \( \text{TP}_i \) and \( \text{FP}_i \) are the true and false positives for class \( i \), respectively, and \( K \) is the number of classes.
(2) \textbf{Recall:} Represents the proportion of actual malware apps correctly classified:
    $\text{Recall} = \frac{\sum_{i=1}^{K} \text{TP}_i}{\sum_{i=1}^{K} (\text{TP}_i + \text{FN}_i)}$,  
    where \( \text{FN}_i \) is the false negatives for class \( i \).
(3) \textbf{Accuracy:} Defines the total number of correct predictions out of all predictions:
    $\text{Accuracy} = \frac{\sum_{i=1}^{n} \mathbf{1}(y_i = \hat{y}_i)}{n}$,  
    where \( y_i \) and \( \hat{y}_i \) are the true and predicted labels, and \( n \) is the total number of samples.
(4) \textbf{F1 Score:} The harmonic mean of precision and recall:
    $\text{F1 score} = \frac{2 \cdot \text{Precision} \times \text{Recall}}{\text{Precision} + \text{Recall}}$.

\BfPara{Kolmogorov-Smirnov (KS) Test} Specific to concept drift, we use the KS test, a non-parametric statistical test used to compare the distributions of two datasets and determine whether they originate from the same underlying probability distribution. KS is particularly useful in detecting concept drift by measuring discrepancies between two empirical distributions. Given two cumulative distribution functions (CDFs), \( F_1(X) \) and \( F_2(X) \), the KS statistic is defined as $D_{n,m} = \sup_x | F_1(x) - F_2(x) |$ where \( D_{n,m} \) is the KS statistic (maximum difference between the two distributions), \( F_1(x) \) is the empirical CDF of the first dataset (e.g., past data), \( F_2(x) \) is the empirical CDF of the second dataset (e.g., new data), and \( \sup_x \) denotes the supremum (i.e., the greatest absolute difference over all \( x \)).
 For two samples of sizes \( n \) and \( m \), the null hypothesis \( H_0 \) states that both datasets come from the same distribution. The critical value for rejecting \( H_0 \) at a given significance level \( \alpha \) is given by $D_{\alpha} = c(\alpha) \sqrt{\frac{n + m}{n m}}$, where \( c(\alpha) \) is a constant determined by the significance level \( \alpha \), and obtained from statistical tables.

\subsection{Experiment Setup}\label{sec:ExperimentalSetup}
\BfPara{Environment} All experiments were conducted using Google Colaboratory (``Colab''), a cloud-based platform that provides free access to computing resources, including GPU acceleration. Colab supports writing and executing Python code directly in a web browser. The experiments were run using both the cloud-hosted and local runtimes, with the default Colab configuration and no manual modifications. The \textit{scikit-learn} library was used to build machine learning models, generate classification reports, and create confusion matrices. For deep learning models, we used \textit{Keras}, a high-level neural network API running on top of TensorFlow.

\BfPara{Experiments Design} 
Each model was evaluated using different subsets of features selected from the Kronodroid dataset. These subsets included all permissions (a total of 166), as well as subsets that excluded permissions based on their protection levels--specifically deprecated, restricted, and not-for-use-by-third-party permissions.

\BfPara{Parameters} The default configurations were used for RF, with the \textit{random state} parameter set to 42. For deep learning algorithms, we used binary cross-entropy as the loss function, Adam optimizer with a default learning rate of $0.001$, an early stopping criterion based on minimum validation loss, 15 training epochs, a batch size of 15, and a validation split of $0.10$. Accuracy and F1 score were used as performance metrics to evaluate the generated models.

\BfPara{Architecture} We use these deep learning architectures: 
\begin{itemize}[leftmargin=*]
    \item \BfPara{CNN} Sequential input shape and two convolutional layers with 64 and 128 filters, each followed by ReLU activation and max pooling layers. This is followed by a dense layer with 128 units, also using ReLU activation, and a dropout layer with a dropout rate of 0.2. The network concludes with a final output layer that features a single sigmoid unit for binary classification.

    \item \BfPara{RNN}  The input data was sequential, and the model included a single RNN layer with 8 units, followed by a flattening layer. A dense layer with 128 units and a dropout layer with a dropout rate of 0.2 were then added, leading to a final output layer with a single sigmoid unit. ReLU activation was used for the hidden layer, and sigmoid activation for the output.

   \end{itemize}

\vspace{2mm}   

\section{Results and Discussion}\label{sec:Experiments}

\subsection{Ignoring the Chronological Context}
In these experiments, we ignored temporal factors and trained and tested on the entire dataset, following a common practice in the literature. We use these results as a baseline to underscore the importance of correctly accounting for concept drift. For machine learning models, the dataset was split into $80\%$ for training and $20\%$ for testing. The same split was applied to the deep learning models, with an additional $10\%$ of the training set reserved for validation.

The results in \autoref{tab:CombinedRealEmulator} compare the performance of RF, CNN, and RNN on data collected from both a real device and an emulator. These classifiers were evaluated using 166 permissions as features, with precision, recall, and F1 score reported separately for malware and benign samples, and accuracy used as an overall metric. Although both datasets (real and emulator) used the same set of permissions, there were minor differences in precision, recall, and F1 scores. RF consistently achieved the best overall performance across both datasets, while CNN and RNN showed slightly lower performance, particularly on the emulator dataset, where they exhibited reduced recall for malware samples.

\begin{table*}[t]
    \centering
    \caption{Performance of various models on real devices and emulators. (All) All permissions: (E) Exclude, (D) deprecated, (R) Restricted, (N) Not for use by third party.}
    \label{tab:CombinedRealEmulator}\vspace{-3mm}
    \begin{tabular}{c|ccc|ccc|c||ccc|ccc|c}
        \hline
        \multirow{3}{*}{\textbf{Model}} & \multicolumn{7}{c||}{\textbf{Real Device}} & \multicolumn{7}{c}{\textbf{Emulator}} \\ 
        \cline{2-15}
        & \multicolumn{3}{c|}{Benign} & \multicolumn{3}{c|}{Malware} & \multirow{2}{*}{\textbf{Acc}} 
        & \multicolumn{3}{c|}{Benign} & \multicolumn{3}{c|}{Malware} & \multirow{2}{*}{\textbf{Acc}} \\ 
        \cline{2-7} \cline{9-14}
        & \textbf{P} & \textbf{R} & \textbf{F1} & \textbf{P} & \textbf{R} & \textbf{F1} &  
        & \textbf{P} & \textbf{R} & \textbf{F1} & \textbf{P} & \textbf{R} & \textbf{F1} &  \\
        \hline
        {RF-All} & 0.943 & 0.959 & 0.951 & 0.964 & 0.949 & 0.956 & 0.954  
                 & 0.942 & 0.965 & 0.953 & 0.956 & 0.927 & 0.941 & 0.948 \\  
        {RF-ED}  & 0.938 & 0.956 & 0.947 & 0.960 & 0.943 & 0.951 & 0.949  
                 & 0.939 & 0.963 & 0.951 & 0.954 & 0.924 & 0.939 & 0.946 \\  
        {RF-ER}  & 0.919 & 0.947 & 0.933 & 0.952 & 0.925 & 0.938 & 0.936  
                 & 0.924 & 0.954 & 0.939 & 0.943 & 0.905 & 0.924 & 0.932 \\  
        {RF-EN}  & 0.937 & 0.957 & 0.947 & 0.960 & 0.943 & 0.951 & 0.949  
                 & 0.939 & 0.962 & 0.950 & 0.953 & 0.925 & 0.939 & 0.945 \\  
        \hline
        {CNN-All} & 0.933 & 0.949 & 0.941 & 0.954 & 0.939 & 0.947 & 0.944  
                 & 0.939 & 0.964 & 0.951 & 0.954 & 0.924 & 0.939 & 0.946 \\  
        {CNN-ED}  & 0.936 & 0.953 & 0.945 & 0.957 & 0.942 & 0.949 & 0.947  
                 & 0.951 & 0.941 & 0.946 & 0.929 & 0.941 & 0.935 & 0.941 \\  
        {CNN-ER}  & 0.913 & 0.944 & 0.928 & 0.948 & 0.919 & 0.933 & 0.931  
                 & 0.923 & 0.934 & 0.928 & 0.919 & 0.905 & 0.912 & 0.921 \\  
        {CNN-EN}  & 0.938 & 0.948 & 0.943 & 0.953 & 0.943 & 0.948 & 0.946  
                 & 0.936 & 0.957 & 0.946 & 0.946 & 0.921 & 0.934 & 0.941 \\  
        \hline
        {RNN-All} & 0.914 & 0.968 & 0.940 & 0.971 & 0.920 & 0.944 & 0.943  
                 & 0.935 & 0.960 & 0.947 & 0.949 & 0.917 & 0.933 & 0.941 \\  
        {RNN-ED}  & 0.931 & 0.941 & 0.936 & 0.948 & 0.939 & 0.944 & 0.940  
                 & 0.937 & 0.953 & 0.945 & 0.942 & 0.923 & 0.932 & 0.939 \\  
        {RNN-ER}  & 0.929 & 0.906 & 0.917 & 0.920 & 0.940 & 0.930 & 0.924  
                 & 0.927 & 0.934 & 0.931 & 0.920 & 0.912 & 0.916 & 0.924 \\  
        {RNN-EN}  & 0.928 & 0.948 & 0.938 & 0.953 & 0.936 & 0.944 & 0.941  
                 & 0.930 & 0.956 & 0.943 & 0.946 & 0.914 & 0.930 & 0.937 \\  
        \hline
    \end{tabular}\vspace{-5mm}
\end{table*}

\takeaway{While static permissions are reliable, the source of the dataset can still influence the detection outcomes.}

For the exclusion strategy, the permission protection levels ({\em deprecated}, {\em restricted}, and {\em not used by a third party}) slightly affected the performance of detection models differently. For \textbf{deprecated permissions (ED)}, their exclusion has minimal effects on RF models, with the accuracy remaining stable (e.g., a slight drop from 0.954 to 0.949 in the real dataset and from 0.948 to 0.946 in the emulator dataset). While RNN results have a similar drop in accuracy as RF, CNN accuracy increased from 0.944 to 0.947 in the real dataset.

The exclusion of the \textbf{restricted permissions (ER)} results in a significant decrease in performance across all models and datasets. In the real dataset, the RF accuracy dropped from 0.954 to 0.933, CNN from 0.947 to 0.933, and RNN from 0.940 to 0.924. The decline is even more pronounced in the emulator dataset, where RF accuracy decreased from 0.948 to 0.932, CNN from 0.944 to 0.931, and RNN from 0.943 to 0.924--a similar effect is observed with the emulator. These results highlight that the restricted permissions contain essential features that boost the models' accuracy. 

For \textbf{not-for-third-party permissions (EN)}, their exclusion has minimal effects across models. In the real dataset, the RF and RNN accuracy remained nearly unchanged (RF: 0.954 to 0.949 and RNN: 0.943 to 0.941), while CNN improved from 0.944 to 0.946. In the emulator dataset, the RF accuracy decreased slightly from 0.948 to 0.945, while the CNN and RNN accuracy experienced a marginal decrease from 0.946 to 0.944 and from 0.941 to 0.937, respectively. This indicates that excluding not-for-third-party permissions has a limited role in influencing model performance.

\takeaway{\small Restricted permissions have a significant impact on model performance, with their exclusion causing notable drops in accuracy and F1 scores across all models and datasets---indicating their importance in distinguishing benign from malicious behavior. In contrast, removing deprecated or third-party-only permissions has minimal effect, suggesting they contribute redundant or less relevant features.}

\subsection{Year-to-Year Strategy}
\label{Corss_Years_Strategy}

\subsubsection{Performance of Models Without KS Test.}
To investigate factors of concept drift that affect the performance of the models, we define the following criteria:

\BfPara{\ding{172} Accuracy and F1 score drift} Based on accuracy and F1 score for all models that ignored the temporal factor, we calculated their averages for both accuracy and F1 score with all permissions, as presented in Tables~\ref{tab:CombinedRealEmulator} and considered as baseline. Any value greater than 0.50 and less than 0.94 is considered to exhibit \textit{drift} in the testing year and is highlighted in red on the heatmap figures.

\BfPara{\ding{173} For balancing results} To evaluate the effect of balance on the performance of the models, we considered any value less than 0.50 to require improvement (\textit{poor performance}). These values are highlighted in yellow to monitor enhancements in the results after balancing.

\BfPara{\ding{174} Outperforming the baseline} Any result greater than 0.94 is highlighted in green, indicating that it \textit{exceeds} the performance of the baseline.

\begin{figure*}[h]
    \centering
    \begin{subfigure}[t]{0.35\textwidth} 
        \centering
        \includegraphics[width=0.90\textwidth]{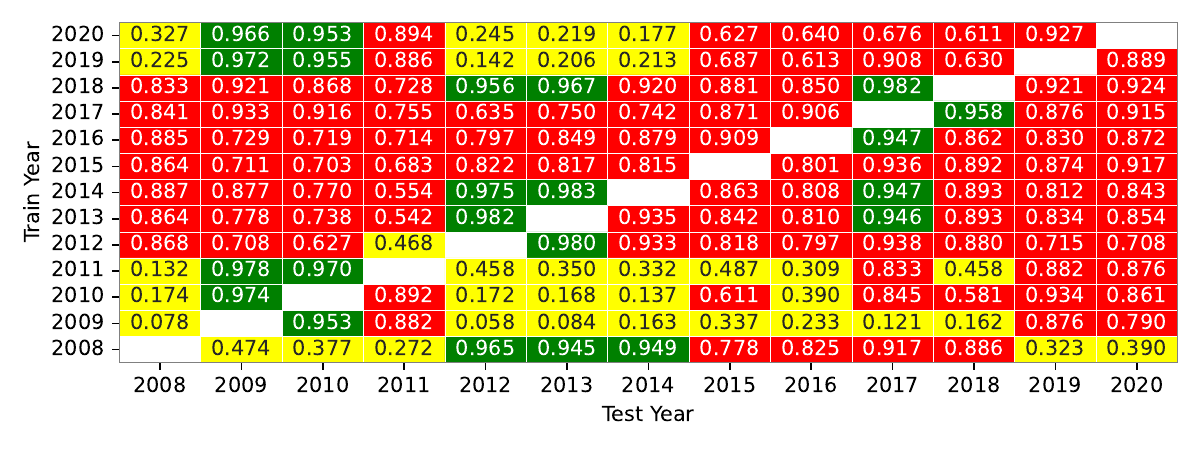}\vspace{-2mm}
        \caption{\normalfont Pre-RF-R-Acc.}
    \label{fig:/Heatmap_Accuracy_Pre_Permissions_ATAT_RF_Real}
    \end{subfigure} 
    ~\hspace{-3em}
    \begin{subfigure}[t]{0.35\textwidth}  
        \centering
        \includegraphics[width=0.90\textwidth]{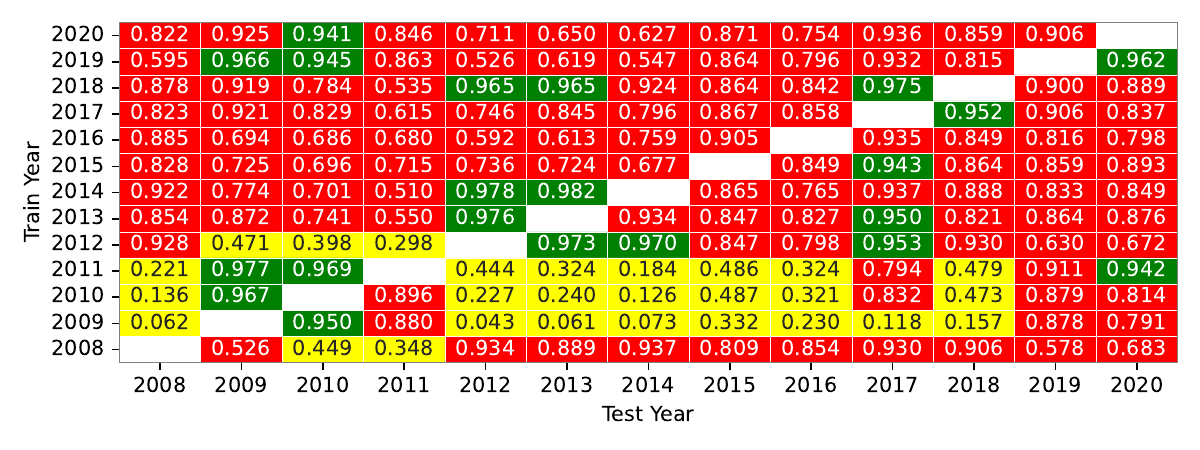}\vspace{-2mm}
        \caption{\normalfont Pre-CNN-R-Acc.}
        \label{Heatmap_Accuracy_Pre_Permissions_ATAT_CNN_Real}
    \end{subfigure}
    ~\hspace{-3em}
    \begin{subfigure}[t]{0.35\textwidth}
        \centering
        \includegraphics[width=0.90\textwidth]{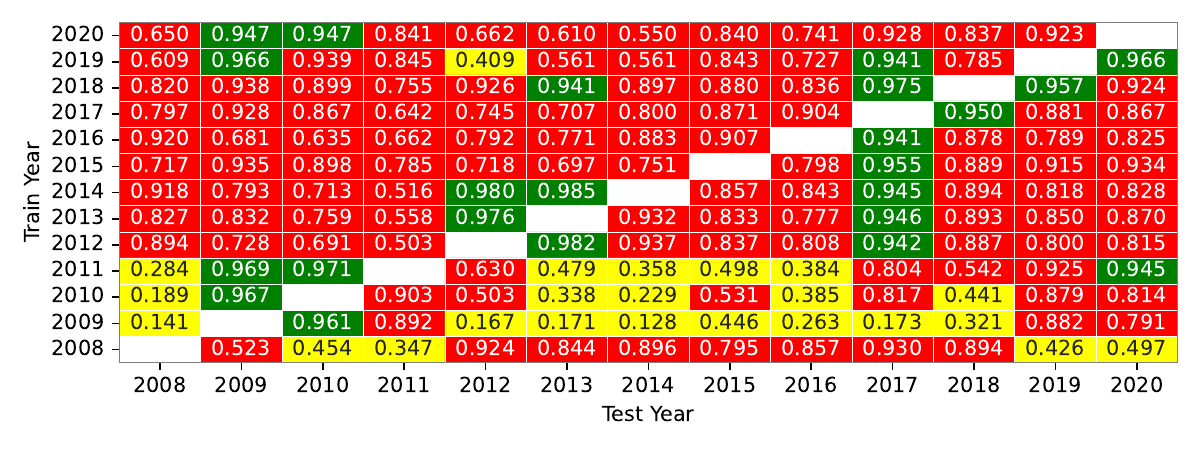}\vspace{-2mm}
        \caption{\normalfont Pre-RNN-R-Acc.}
        \label{Heatmap_Accuracy_Pre_Permissions_ATAT_RNN_Real}
    \end{subfigure}

    \begin{subfigure}[t]{0.35\textwidth} 
        \centering
        \includegraphics[width=0.90\textwidth]{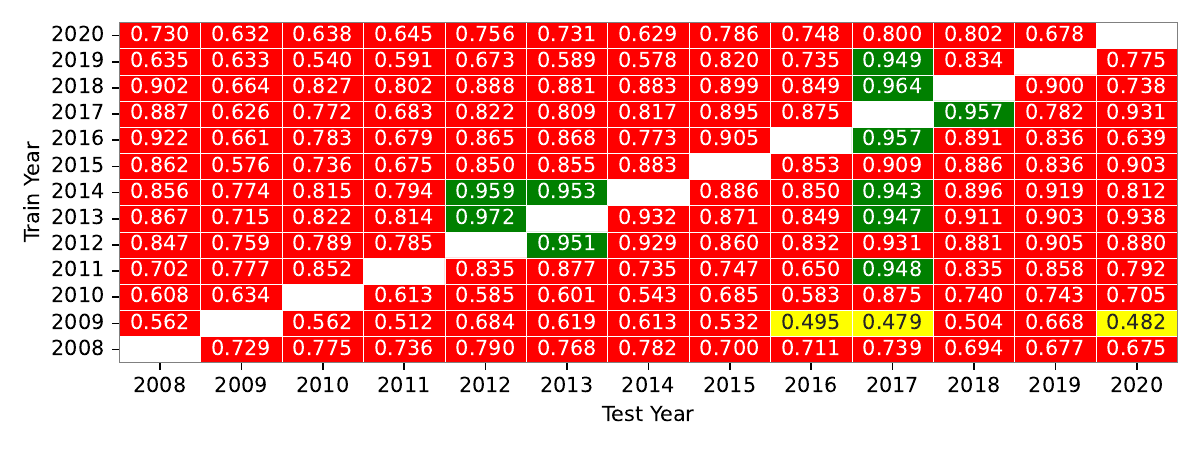}\vspace{-2mm}
        \caption{\normalfont Post-RF-R-Acc.}
    \label{fig:/Heatmap_Accuracy_Post_Permissions_ATAT_RF_Real}
    \end{subfigure} 
    ~\hspace{-3em}
    \begin{subfigure}[t]{0.35\textwidth}  
        \centering
        \includegraphics[width=0.90\textwidth]{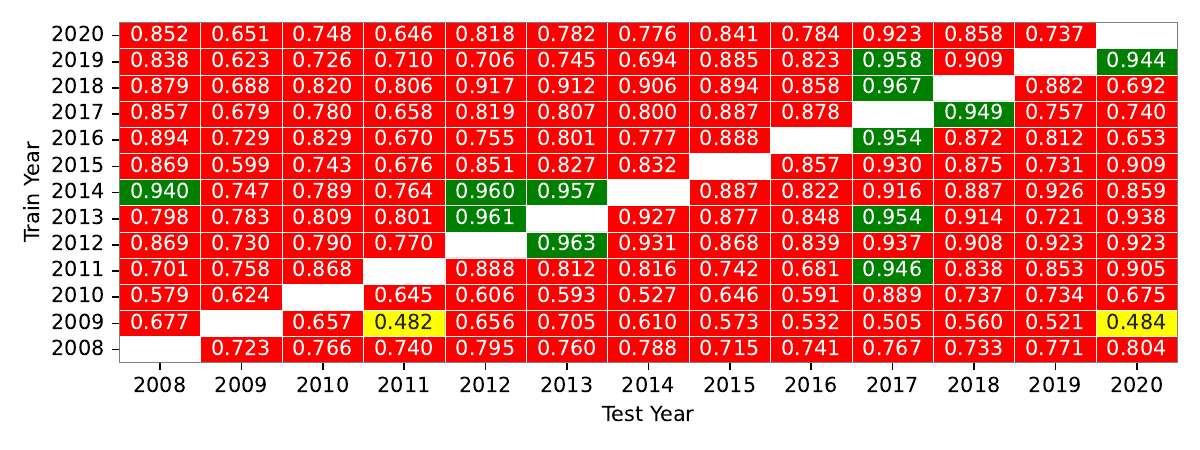}\vspace{-2mm}
        \caption{\normalfont Post-CNN-R-Acc.}
        \label{Heatmap_Accuracy_Post_Permissions_ATAT_CNN_Real}
    \end{subfigure}
    ~\hspace{-3em}
    \begin{subfigure}[t]{0.35\textwidth}
        \centering
        \includegraphics[width=0.90\textwidth]{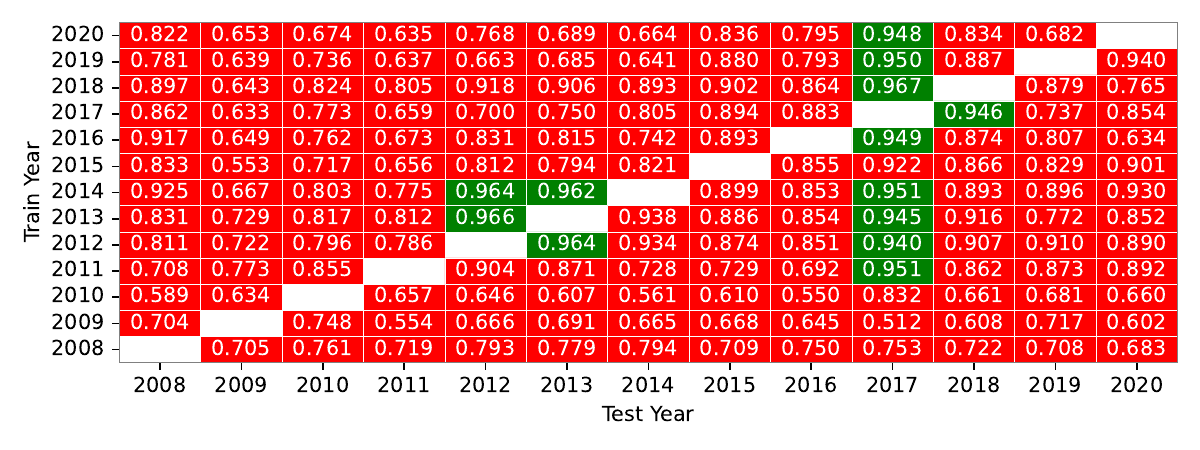}\vspace{-2mm}
        \caption{\normalfont Post-RNN-R-Acc.}
        \label{Heatmap_Accuracy_Post_Permissions_ATAT_RNN_Real}
    \end{subfigure}

\vspace{-3mm}
    \caption{Accuracy performance of various models pre/post balancing with all permissions. (R) real device dataset. Cells with red color indicate decreased accuracy, yellow represent values less than 0.50, and green show improvements.}
    \label{fig:Heatmap_Acc_all_Pre_Post}\vspace{-2mm}
\end{figure*}

\BfPara{Heatmaps For Emulator and Real Datasets}
For each model, we generated heatmaps for the real and emulator data, both before and after balancing, as shown in \autoref{fig:Heatmap_Acc_all_Pre_Post}. In addition, we summarize these heatmap results in tables for each model (RF, CNN, and RNN) presented in~\autoref{tab:StackedSummaryAll}. The main observations of the results are as follows. 

\BfPara{\ding{172} Pre-balancing} RF models on both real and emulator datasets show a significant number of drift accuracy in heatmap cells (100+ in most cases), highlighting substantial concept drift in accuracy over many testing years. However, RF demonstrates slightly less drift in accuracy compared to CNN (e.g., 108 for the real dataset and 111 for the emulator dataset). RNN follows a similar trend, and RF generally has fewer drifts than RNN. In contrast, the count of poor accuracy results, representing values below 0.50, is relatively high, particularly for the emulator dataset, which reached 47 with RF after excluding restricted permissions, while CNN and RNN achieved 28 and 29, respectively. For exceeding baseline accuracy, the models achieved more than 20 results in some cases, but the impact of balancing significantly reduced this number by approximately half, revealing the strong influence of class imbalance on performance. 
Excluding D and R features reduced the drift in accuracy for RF compared to using all permissions, while a decrease was observed in CNN for both real and emulator datasets, and only in the real dataset for RNN. In contrast, excluding N permissions reduced the accuracy drift only in the RNN. 

\takeaway{\small Excluding D, R, and N permissions generally improves performance stability across years. However, this can come at the cost of reduced accuracy, particularly on emulator datasets, highlighting a trade-off between stability and performance. Moreover, fewer instances surpass baseline performance after exclusion, suggesting limited gains in certain scenarios.}

\BfPara{\ding{173} Post-Balancing}
The results of the models' performance post-balancing show that balancing the datasets positively impacts the F1 scores across all models (RF, CNN, and RNN). This improvement highlights that addressing the imbalance of data before analyzing the drift is crucial to accurately assess the behavior of the models. The increase in F1 scores indicates that the models benefit from a more representative distribution of benign and malicious samples, which enhances their ability to generalize over testing years. Moreover, the reduction in poor performance in accuracy is clear, which indicates that model performance improved when compared with pre-balance. However, the increase in drift count in accuracy, especially for CNN and RNN, reflects the fundamental reality of concept drift. This indicates that balancing the dataset improves model performance and reveals the extent of concept drift previously obscured by imbalance. Once this imbalance is addressed, the models' results reveal more accurate trends, highlighting the impact of drift over time.
\takeaway{\small Balancing the datasets before addressing the concept drift is critical to revealing the drift's extent and mitigating its impact.}

Focusing on the effect of excluding D, R, and N permissions, the RF results demonstrate resilience to concept drift, with fewer increases in the number of drifts compared to CNN and RNN. Excluding D for RF with real is not affected, and concept drift occurred with the emulator in one new year, similar to excluding R (i.e., an increase from 145 to 146 poor performance instances). However, the drift is reduced when N is excluded--by one year in real and two in the emulator. CNN is the most sensitive to the exclusion of permissions, specifically D and R, where the poor accuracy values increase significantly. RNN shows a moderate response to the exclusion scenarios. Excluding D and R permissions increases concept drift, especially in emulator datasets. However, RNN still achieves improvements in the post-balance F1 scores for the real dataset with all permissions. Additionally, deep learning models show less poor accuracy and higher F1 scores after balancing.
\takeaway{\small Although deprecated or restricted permissions may seem outdated, they remain relevant for many applications and aid malware detection. Results show that excluding them increases concept drift in CNN and RNN models, highlighting their continued real-world predictive value.}

\begin{table*}[t]
\centering
\caption{RF, CNN, and RNN accuracy and F1 scores for different configurations of permission sets with a comparison to the baseline. Red indicates the count of testing years with decreased accuracy, yellow represents values less than 0.50, and green shows the number of years improved. (D) Deprecated, (R) Restricted, and (N) Not for use by third party.}
\label{tab:StackedSummaryAll}\vspace{-3mm}
\scalebox{0.8}{\begin{tabular}{l|ccc|ccc}
\hline
\multirow{2}{*}{\textbf{Balance}} & \multicolumn{3}{c|}{\textbf{Accuracy}} & \multicolumn{3}{c}{\textbf{F1 score}} \\  
\cline{2-7}
& \textbf{Red} & \textbf{Yellow} & \textbf{Green} & \textbf{Red} & \textbf{Yellow} & \textbf{Green} \\  
\hline
\multicolumn{7}{c}{\textbf{RF - Random Forest}} \\  
\hline
Pre  & 100 & 34 & 22  & 111 & 44 & 1 \\  
Pre (Excl.) & 99 (D) / 95 (R) / 101 (N)  & 35 (D) / 40 (R) / 34 (N)  & 22 (D) / 21 (R) / 21 (N)  
            & 109 (D) / 104 (R) / 112 (N)  & 45 (D) / 51 (R) / 42 (N)  & 2 (D) / 1 (R) / 2 (N) \\  
Post  & 142 & 3 & 11  & 136 & 9 & 11 \\  
Post (Excl.) & 142 (D) / 146 (R) / 141 (N)  & 4 (D) / 4 (R) / 5 (N)  & 10 (D) / 6 (R) / 10 (N)  
             & 135 (D) / 136 (R) / 135 (N)  & 11 (D) / 14 (R) / 11 (N)  & 10 (D) / 6 (R) / 10 (N) \\  
\hline
\multicolumn{7}{c}{\textbf{CNN - Convolutional Neural Network}} \\  
\hline
Pre  & 108 & 27 & 21  & 111 & 44 & 1 \\  
Pre (Excl.) & 112 (D) / 109 (R) / 110 (N)  & 24 (D) / 29 (R) / 23 (N)  & 20 (D) / 18 (R) / 23 (N)  
            & 110 (D) / 113 (R) / 120 (N)  & 44 (D) / 42 (R) / 34 (N)  & 2 (D) / 1 (R) / 2 (N) \\  
Post  & 142 & 2 & 12  & 136 & 8 & 12 \\  
Post (Excl.) & 146 (D) / 146 (R) / 143 (N)  & 0 (D) / 1 (R) / 1 (N)  & 10 (D) / 9 (R) / 12 (N)  
             & 142 (D) / 140 (R) / 139 (N)  & 4 (D) / 7 (R) / 5 (N)  & 10 (D) / 9 (R) / 12 (N) \\  
\hline
\multicolumn{7}{c}{\textbf{RNN - Recurrent Neural Network}} \\  
\hline
Pre  & 110 & 23 & 23  & 116 & 38 & 2 \\  
Pre (Excl.) & 107 (D) / 108 (R) / 108 (N)  & 28 (D) / 27 (R) / 26 (N)  & 21 (D) / 21 (R) / 22 (N)  
            & 117 (D) / 112 (R) / 114 (N)  & 38 (D) / 42 (R) / 40 (N)  & 1 (D) / 2 (R) / 2 (N) \\  
Post  & 143 & 0 & 13  & 140 & 3 & 13 \\  
Post (Excl.) & 142 (D) / 147 (R) / 144 (N)  & 1 (D) / 2 (R) / 0 (N)  & 13 (D) / 7 (R) / 12 (N)  
             & 140 (D) / 141 (R) / 139 (N)  & 3 (D) / 8 (R) / 5 (N)  & 13 (D) / 7 (R) / 12 (N) \\  
\hline
\end{tabular}}
\end{table*}

\subsubsection{Concept Drift Detection Using KS Test}
    In this set of experiments, we aim to detect the presence of concept drift using the Kolmogorov-Smirnov (KS) test. We employ the KS test to analyze feature distributions across different time periods, datasets (real vs. emulator), and permission configurations (e.g., exclusion of deprecated, restricted, or not-for-use-by-third-party permissions).

    The null hypothesis ($H_0$) proposes that there is no significant difference in the accuracy and F1 score distributions between the two years being compared. If the $p$-value resulting from the KS test is below a chosen significance threshold $(p \leq 0.05)$, we reject the null hypothesis, indicating that a statistically significant concept drift has occurred.

\begin{figure*}[t]
    \centering

     %------------------------ second  Acc Emu -------------------------------
      \begin{subfigure}[t]{0.35\textwidth} 
        \centering
        \includegraphics[width=0.90\textwidth]{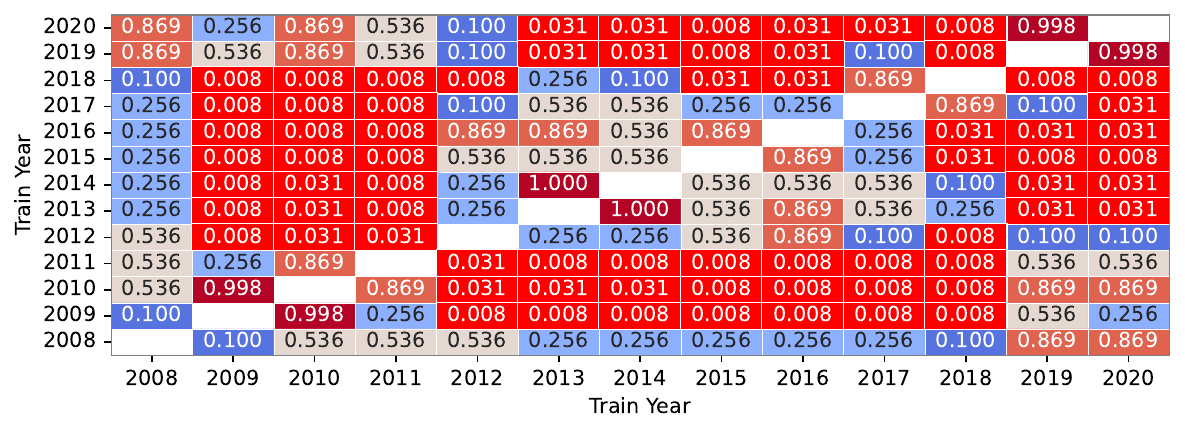}\vspace{-2mm}
        \caption{\normalfont Pre-RF-E-Acc.}
    \label{fig:/KSHeatmap_Accuracy_Pre_Permissions_ATAT_RF_Emulator}
    \end{subfigure} 
    ~\hspace{-3em}
    \begin{subfigure}[t]{0.35\textwidth}  
        \centering
        \includegraphics[width=0.90\textwidth]{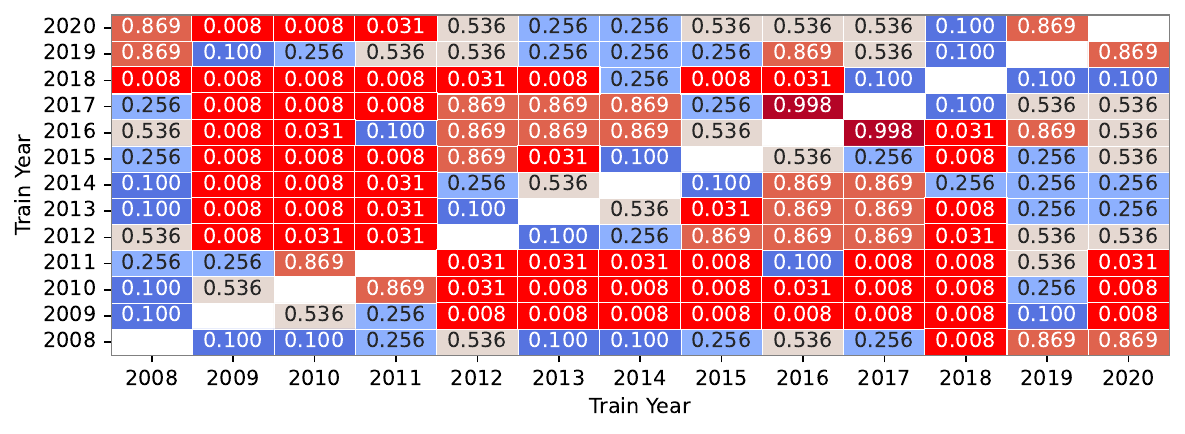}\vspace{-2mm}
        \caption{\normalfont Pre-CNN-E-Acc.}
        \label{KSHeatmap_Accuracy_Pre_Permissions_ATAT_CNN_Emulator}
    \end{subfigure}
    ~\hspace{-3em}
    \begin{subfigure}[t]{0.35\textwidth}
        \centering
        \includegraphics[width=0.90\textwidth]{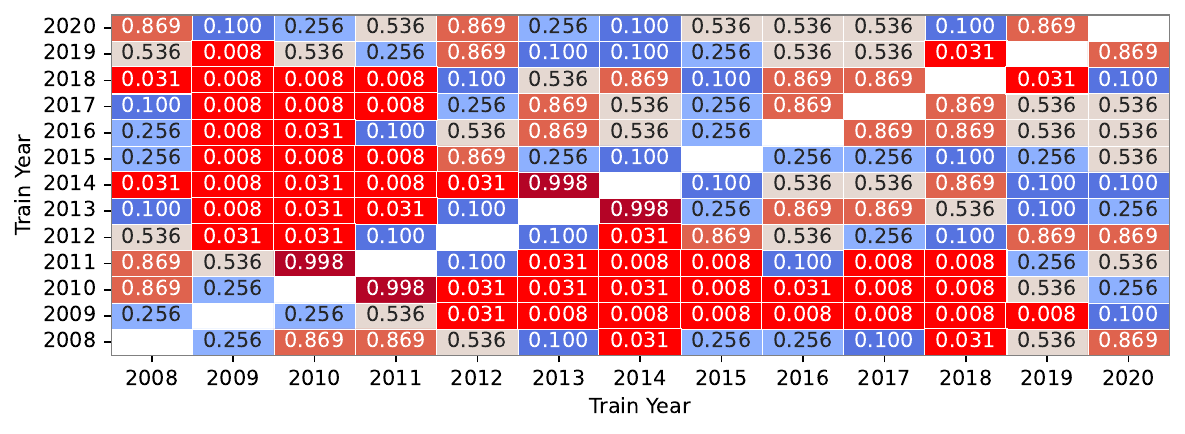}\vspace{-2mm}
        \caption{\normalfont Pre-RNN-E-Acc.}
        \label{KSHeatmap_Accuracy_Pre_Permissions_ATAT_RNN_Emulator}
    \end{subfigure}
    ~\hspace{-2.2em}
     
     %------------------------ second  Acc Emu -------------------------------
      \begin{subfigure}[t]{0.35\textwidth} 
        \centering
        \includegraphics[width=0.90\textwidth]{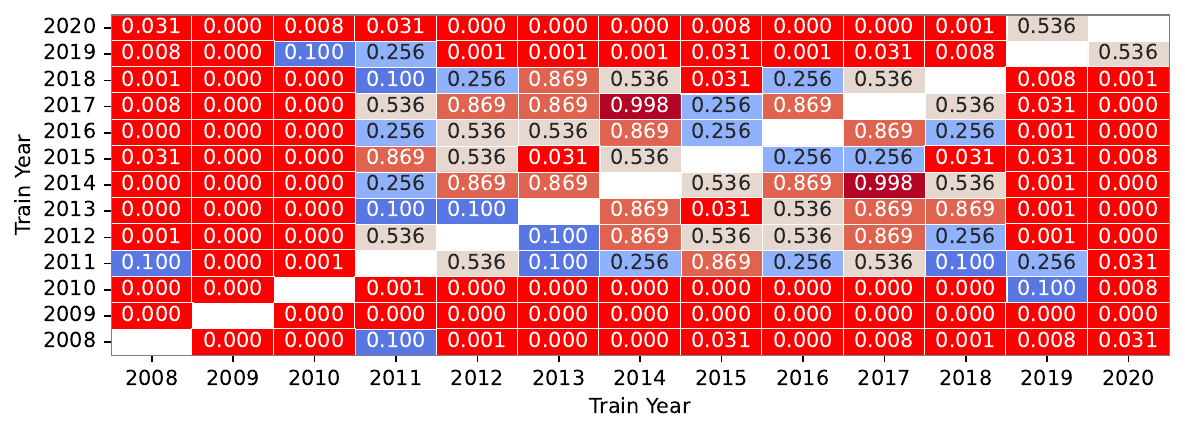}\vspace{-2mm}
        \caption{\normalfont Post-RF-E-Acc.}
    \label{fig:/KSHeatmap_Accuracy_Post_Permissions_ATAT_RF_Emulator}
    \end{subfigure} 
    ~\hspace{-3em}
    \begin{subfigure}[t]{0.35\textwidth}  
        \centering
        \includegraphics[width=0.90\textwidth]{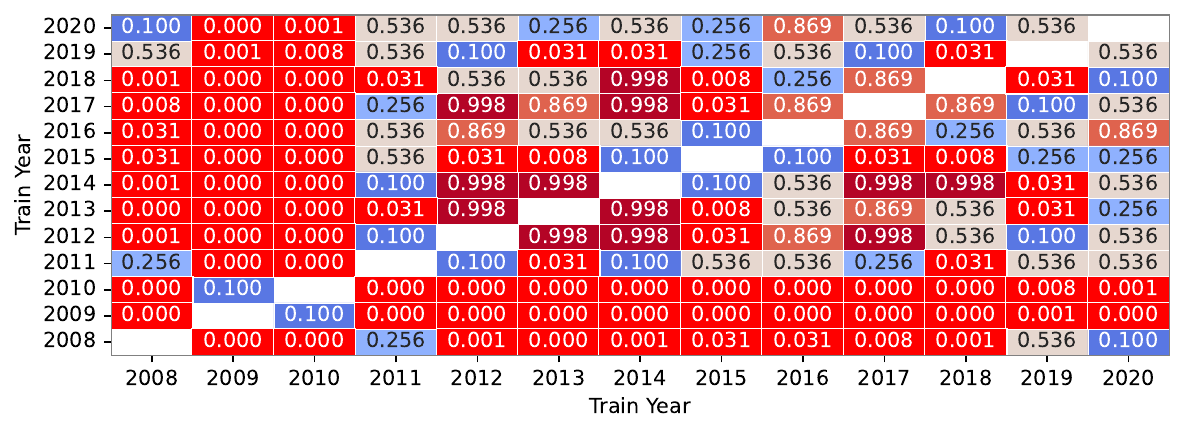}\vspace{-2mm}
        \caption{\normalfont Post-CNN-E-Acc.}
        \label{KSHeatmap_Accuracy_Post_Permissions_ATAT_CNN_Emulator}
    \end{subfigure}
    ~\hspace{-3em}
    \begin{subfigure}[t]{0.35\textwidth}
        \centering
        \includegraphics[width=0.90\textwidth]{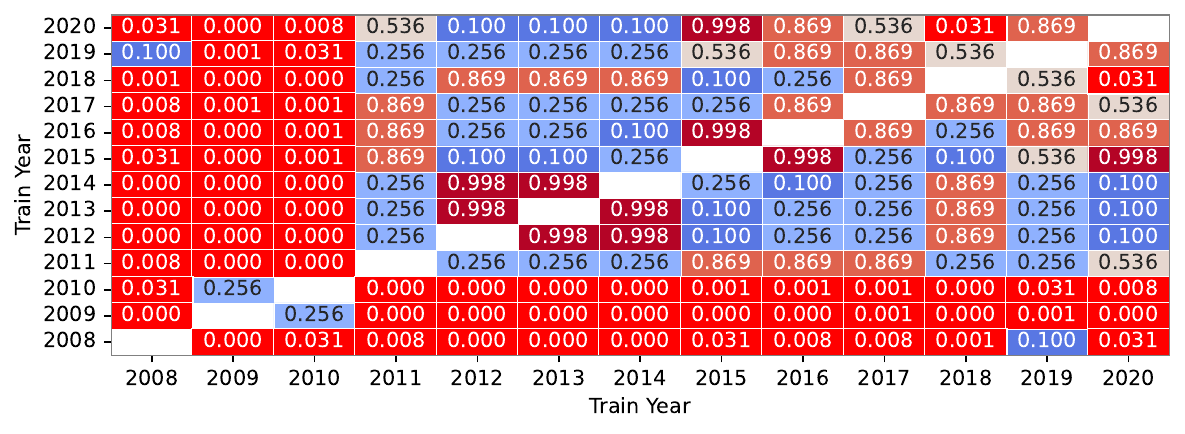}\vspace{-2mm}
        \caption{\normalfont Post-RNN-E-Acc.}
        \label{KSHeatmap_Accuracy_Post_Permissions_ATAT_RNN_Emulator}
    \end{subfigure}
    ~\hspace{-2.2em}
    \vspace{-3mm}
    \caption{Concept drift detection of various algorithms pre/post balancing, where $p$-value $\leq 0.05$ (red color). (E) emulator}
    \label{fig:KSHeatmap_Acc_all_Pre_Post}\vspace{-2mm}
\end{figure*}

    We conducted a large number of experiments, totaling 32 heatmap results for each model and~\autoref{fig:KSHeatmap_Acc_all_Pre_Post} illustrates an example of the results obtained for the three classification models (RF, CNN, and RNN) before and after balancing, using both the real and emulator datasets. Each heatmap represents the $p$-values of the KS tests applied to the accuracy and F1 score distributions for year-to-year comparisons under various permission configurations.% (e.g., excluding deprecated, restricted, or third-party permissions).

    To simplify interpretation, we summarize each heatmap by counting the number of year-to-year comparisons with a $p$-value less than 0.05, indicating a statistically significant concept drift. These summarized results are presented separately for each model in~\autoref{tab:StackedKSSummaryAll}.

\BfPara{\ding{172} Impact of Balancing}
Across all models, the number of significant $p$-values noticeably increased after balancing, indicating that balancing better captures concept drift in both accuracy and F1 scores. This suggests that balancing has a significant impact on improving the sensitivity of drift detection. For example, in RF, the number of significant $p$-values for accuracy increased from 52 to 84 (real, all permissions) and from 70 to 96 (emulator, all permissions) after balancing. For CNN, a similar trend is observed, with $p$-values increasing from 52 to 60 and from 58 to 76. For RNN, the increase was from 40 to 60 and from 48 to 64.

\takeaway{\small Balancing the dataset not only improves concept drift detection in general but also reveals variations in model sensitivity to concept drift.}

\BfPara{\ding{173} Dataset-Specific Observations (Real vs. Emulator)}
For all models, the $p$-value counts are generally higher when using the emulator dataset compared to the real dataset. For the RF model, the significant $p$-values related to post-balancing accuracy increased to 96 for the emulator dataset, compared to 84 for the real dataset. Similarly, for the CNN and RNN models, the $p$-values for accuracy reached 76 compared to 60, and 64 compared to 60, respectively.

\begin{table*}[t]
\centering
\caption{RF, CNN, and RNN model $p$-value counts ($\leq 0.05$) for accuracy and F1 score across real devices and emulators. (Pre) Pre-balanced, (Post) Post-balanced, (All) All permissions, (D) Deprecated, (R) Restricted, (N) Not for use by 3rd-party.}
\label{tab:StackedKSSummaryAll}\vspace{-3mm}
\begin{tabular}{l|cc|cc||cc|cc}
\hline
\multirow{3}{*}{\textbf{Balance}} & \multicolumn{4}{c||}{\textbf{Real Device}} & \multicolumn{4}{c}{\textbf{Emulator}} \\  
\cline{2-9}
& \multicolumn{2}{c|}{Accuracy ($p \leq 0.05$)} & \multicolumn{2}{c||}{F1 score ($p \leq 0.05$)} & \multicolumn{2}{c|}{Accuracy ($p \leq 0.05$)} & \multicolumn{2}{c}{F1 score ($p \leq 0.05$)} \\ 
\cline{2-9}
& All & Excl. & All & Excl. & All & Excl. & All & Excl. \\ 
\hline
\multicolumn{9}{c}{\textbf{RF - Random Forest}} \\  
\hline
Pre  & 52 & 58 (D) / 42 (R) / 58 (N)  & 50 & 64 (D) / 50 (R) / 60 (N) & 70 & 68 (D) / 58 (R) / 68 (N)  & 66 & 64 (D) / 56 (R) / 58 (N) \\  
Post & 84 & 94 (D) / 92 (R) / 92 (N)  & 80 & 94 (D) / 94 (R) / 92 (N) & 96 & 102 (D) / 96 (R) / 92 (N) & 96 & 102 (D) / 98 (R) / 94 (N) \\  
\hline
\multicolumn{9}{c}{\textbf{CNN - Convolutional Neural Network}} \\  
\hline
Pre  & 52 & 52 (D) / 44 (R) / 52 (N)  & 46 & 38 (D) / 44 (R) / 34 (N) & 58 & 52 (D) / 54 (R) / 58 (N)  & 60 & 62 (D) / 48 (R) / 50 (N) \\  
Post & 60 & 72 (D) / 68 (R) / 58 (N)  & 62 & 76 (D) / 70 (R) / 54 (N) & 76 & 70 (D) / 72 (R) / 64 (N) & 78 & 70 (D) / 72 (R) / 62 (N) \\  
\hline
\multicolumn{9}{c}{\textbf{RNN - Recurrent Neural Network}} \\  
\hline
Pre  & 40 & 36 (D) / 42 (R) / 54 (N)  & 36 & 38 (D) / 40 (R) / 42 (N) & 48 & 42 (D) / 40 (R) / 52 (N)  & 42 & 38 (D) / 46 (R) / 42 (N) \\  
Post & 60 & 72 (D) / 68 (R) / 64 (N)  & 60 & 72 (D) / 68 (R) / 62 (N) & 64 & 60 (D) / 76 (R) / 58 (N) & 64 & 60 (D) / 78 (R) / 58 (N) \\  
\hline
\end{tabular}\vspace{-5mm}
\end{table*}

\BfPara{\ding{174} Impact of Excluding Permissions}
    Across all models, the exclusion of permissions increased concept drift, particularly after balancing. For RF~\ref{tab:StackedKSSummaryAll}, post-balancing accuracy $p$-values for the real dataset increased from 84 (All Permissions) to 94 (Exclude D). Furthermore, CNN and RNN showed improvements, with post-balancing accuracy $p$-values increasing from 60 to 72 for the real dataset.

\begin{figure}[t]
    \centering
    \begin{tikzpicture}
        \begin{axis}[
            ybar,
            width=9cm,
            height=4cm,
            bar width=4pt,
            symbolic x coords={2008,2009,2010,2011,2012,2013,2014,2015,2016,2017,2018,2019,2020},
            xtick=data,
            x tick label style={rotate=45, anchor=east},
            ymin=0, ymax=1,
            xlabel={Train Year},
            ylabel={Average Accuracy},
            enlargelimits=false,
            legend style={at={(.65,0.5)}, anchor=north, legend columns=1,font=\small},
            nodes near coords,  % Adds numbers on top of bars
            every node near coord/.append style={font=\tiny, rotate=90, anchor=west}, % Rotates numbers vertically
            enlarge x limits=0.05
        ]
            % First dataset
            \addplot[ybar, color=blue, pattern=grid] coordinates {(2008,0.732) (2009,0.559) (2010,0.660) (2011,0.801) (2012,0.862) 
                                  (2013,0.878) (2014,0.871) (2015,0.819) (2016,0.815) (2017,0.821) 
                                  (2018,0.850) (2019,0.696) (2020,0.715)};
            
            % Second dataset
            \addplot[ybar, color=blue, pattern=crosshatch] coordinates {(2008,0.727) (2009,0.546) (2010,0.665) (2011,0.836) (2012,0.861) 
                                  (2013,0.879) (2014,0.868) (2015,0.801) (2016,0.846) (2017,0.817) 
                                  (2018,0.849) (2019,0.679) (2020,0.669)};

            \legend{All Permissions, Excluding Deprecated}
        \end{axis}
    \end{tikzpicture}\vspace{-3mm}
    \caption{RF with all permissions and without deprecated.}\label{fig:BarComaper_RF_All_ExcludeD}\vspace{-5mm}
\end{figure}
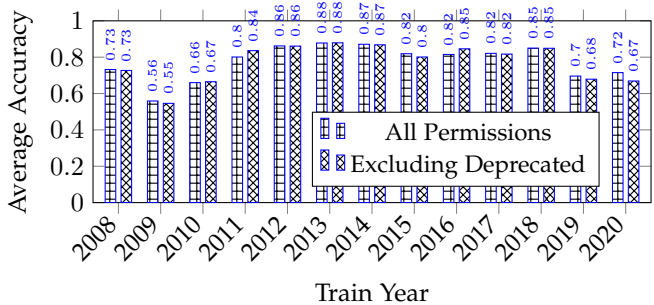

    We further compared the performance of the models with all permissions and after excluding D permissions in~\autoref{fig:BarComaper_RF_All_ExcludeD} for RF with post-balancing on the real dataset. The results of excluding deprecated permissions lead to some, but not significant, effect on model performance across most years. For example, in 2010, the average accuracy improved from 0.660 (all permissions) to 0.665 (D excluded). In 2011, accuracy increased from 0.801 to 0.836, and in 2016, it increased from 0.815 to 0.846. However, in some years, such as 2013, the exclusion has minimal impact, with accuracy remaining nearly the same (0.878 vs. 0.879).

%\BfPara{\ding{71} Takeaway} 
\takeaway{\small While excluding permissions does not have a significant effect on the model's performance, it is statistically significant in improving concept drift detection.}

\subsection{Discussion}
Our results provide strong empirical evidence that concept drift is prevalent in Android malware detection and is influenced by multiple factors. Below, we highlight those results by answering the research questions in section~\ref{sec:Methodology}.

\noindent\textbf{AQ-1:} The findings indicate that permissions play a significant role in both the accuracy and effectiveness of malware detection across all models. Specifically, as shown in the results for the real and emulator datasets in~\autoref{tab:CombinedRealEmulator}, the RF model achieved an accuracy of 0.954 on the real dataset using only 166 permissions out of 489 dynamic and static features. When using these hybrid features, the models achieved accuracies of 0.980 (RF), 0.970 (CNN), and 0.980 (RNN). These results underscore the critical importance of permissions in Android malware detection.

\noindent\textbf{AQ-2:} 
The sensitivity of permissions to concept drift significantly impacts the temporal performance of Android malware detection models. The results reveal that permissions in the emulator dataset are more prone to drift, as evidenced by fluctuations in accuracy and consistently higher $p$-value counts in this dataset compared to the real device dataset. 

Drift became more pronounced after data balancing, as observed across all models. Although accuracy improved in low-performing cases, this improvement also highlights the heightened sensitivity of permissions to distributional changes. For example, accuracy (yellow) increased from 47 to 8 after balancing and excluding restricted permissions in the emulator dataset~\autoref{tab:StackedSummaryAll}. While this is considered a performance improvement, it also coincides with an increase in detected concept drift, which rose from 96 to 146. Additionally, for the same case, the number of significant $p$-values increased from 58 to 96, as shown in~\autoref{tab:StackedKSSummaryAll}.

\noindent\textbf{AQ-3:} 
The exclusion strategy revealed a marginal impact on the models when {\em deprecated} and {\em not for use by third-party} permissions were omitted. In some cases, such as with the CNN model, performance metrics like accuracy and F1 score showed slight improvements, suggesting that these permissions—while historically valuable—may contribute less to a model's ability to generalize in certain scenarios. 

From the perspective of concept drift, the exclusion of these permissions played a significant role in enhancing the detection of drift patterns. Using the KS test or heatmap cells to highlight drift, the results indicated that removing these permissions made concept drift more apparent. This can be attributed to the elimination of outdated or less relevant features, which appeared in certain years but were absent in others, as shown in the analysis section (\autoref{fig:DeprecatedEmulatorReal}).

Thus, while excluding {\em deprecated} and {\em not-for-use-by-third-party} permissions might not always boost immediate model performance, it contributes to the understanding and detection of concept drift--an essential factor for the long-term stability and reliability of malware detection systems.

\section{Related Work}\label{sec:related}

The Android permission system plays a central role in both malware detection and platform security. In the following, we review work on its evolution, abuse, and implications for machine learning-based detection.

\BfPara{Android Permission System}
Since API 6.0, Android's permission model has undergone several changes, expanding to support hardware rather than refining security granularity~\cite{ZhauniarovichG16}. Studies reveal overprivileged apps---especially pre-installed ones---contributing to user confusion and increased risk~\cite{WeiGNF12}. Tools like PScout~\cite{AuZHL12} and Cusper~\cite{TuncayDGG18} exposed structural flaws, while fine-grained enforcement via frameworks like Sorbet~\cite{FragkakiBJS12} and DroidCap~\cite{DawoudB19} sought to address least-privilege enforcement. Recent efforts also uncovered issues with undocumented permissions~\cite{ZhouWWLZC021} and inter-version inconsistencies~\cite{ZhauniarovichG16}, highlighting challenges in sustaining a secure permission framework.

\BfPara{Privilege Escalation via Permission Abuse} The permission system has been exploited through custom permission spoofing~\cite{LiDLYLG22}, exposed components for sensor access~\cite{AldoseriOC22}, covert collusion channels~\cite{ReardonFWOVE19}, and physical hijacking~\cite{WangSCL22}. Tools like PmDroid~\cite{GaoLWS15} detect permission violations by ad libraries, while others reveal permissionless exfiltration paths via shared resources~\cite{WuWYWLW15}. Detection techniques span static (e.g., COVERT~\cite{BagheriSGM15}), dynamic (e.g., VetDroid~\cite{ZhangYYGNZ14}), and hybrid approaches (e.g., Permlyzer~\cite{XuZZ13}), reinforcing the need for multi-pronged analysis and mitigation.

\BfPara{Runtime Permission Issues and Testing}
The shift to runtime permissions introduced new vulnerabilities. Tools such as RevDroid~\cite{FangHLGGWQC16} and Aper~\cite{WangWZWLLC22} analyzed app behavior post-revocation, showing malware often handles revocation more gracefully than benign apps. Testing frameworks like PermDroid~\cite{YangZS22}, PATDroid~\cite{SadeghiJM17}, and DPC~\cite{HsuLHLWC21} automate state-based testing or introduce dynamic controls to restrict third-party library misuse.

% \BfPara{Usability and User Perception of Permissions}
% Studies reveal poor user comprehension of permissions~\cite{PrangeKKDLA24}, common runtime issues~\cite{WangWWLXCYZ23}, and the need for better developer resources~\cite{OishweeSC24,Tuncay24}. User-centric designs, such as flexible models~\cite{ScocciaMASI21} and UI-guided controls~\cite{MalviyaLKTSJ22}, aim to align access control with user expectations. Research also explored tools to generate permission rationale~\cite{WangWTZW20}, leveraging descriptions~\cite{ScocciaPPCK19,QuRZCZC14} and crowdsourced insights~\cite{MomenBF20}.

\BfPara{Android Malware Detection} Permissions are a critical feature in Android malware detection. Numerous works highlight their utility: Ilham~\etal~\cite{IlhamAA18} and Shatnawi~\etal~\cite{ShatnawiYY22} showed high accuracy with selected permissions using traditional classifiers. Sahin~\etal~\cite{SahinKAK23} demonstrated the benefit of reducing permission sets, while Kato~\etal~\cite{KatoSS21} introduced composition ratios to classify malware. Recent models combine permissions with APIs, intents, and behavior traces. Tools such as PermPair~\cite{AroraPC20}, DroidXGB~\cite{KumarS24}, and MalPat~\cite{TaoZGL18} enhance classification by mining permission patterns and graphs. Multi-feature models, including CNNs with opcode and API information~\cite{MillarMRM21}, improve zero-day detection. Federated learning approaches like FEDroid~\cite{FangHLLC0HZ23} address privacy-preserving malware detection using permission-based features. Permissions also support behavior-specific multi-label classification frameworks~\cite{QiaoFCZL22}. These studies reinforce the critical role of permissions in malware detection. However, their effectiveness is susceptible to changes in permission availability and semantics over time, introducing a potential source of concept drift.

\BfPara{Concept Drift} To improve model adaptability under drift, approaches like TRANSCEND~\cite{JordaneySDWPNC17}, TRANSCENDENT~\cite{BarberoPPC22}, and DREAM~\cite{HeLQR24} incorporate conformal prediction, rejection mechanisms, or semi-supervised learning. Active learning and retraining techniques~\cite{MolinaCoronadoMMM23, AbusnainaASAJSM24, AlamFMR24} reduce labeling cost and improve drift responsiveness. Other models use self-training~\cite{AlamFMR24}, contrastive learning~\cite{ChenDWCT23}, and ensemble classifiers with adaptive feature selection~\cite{HuMZLYL17}.

Novel architectures include cluster-based drift detection~\cite{MishraS25}, GNNs for invariant feature learning~\cite{LiIKB25}, and hybrid models enhanced with evolutionary strategies~\cite{GondM25}. Recent insights from Chow~\etal~\cite{ChowKLCAP23} show that drift stems not only from evolving malware but also goodware changes and family-specific shifts. Molina~\etal~\cite{MolinaCoronadoMMM23} further advocate for retraining-on-drift rather than fixed intervals to conserve resources and improve stability.

\BfPara{This Study} While prior work broadly addresses evolving threats and concept drift, our study focuses on the underexplored impact of \textbf{permission deprecation}---a system-level change---on model stability and performance drift. We offer new insights into how Android's evolving permission architecture affects the reliability of ML-based malware detection. Effectively addressing concept drift is critical for sustaining detection accuracy, and this study examines the temporal evolution of permission-based features, demonstrating that even simple models are susceptible to degradation. By isolating the effects of permission deprecation and quantifying drift using accuracy and F1 scores, our findings support the design of more adaptive and resilient detection frameworks.

\section{Conclusion}\label{sec:conclusion}
This work explored the role of Android permissions in malware detection, their sensitivity to concept drift, and the impact of deprecated and restricted permissions on model stability. Using the KronoDroid dataset (including real and emulator devices), it found permissions to be strong features for detection. Excluding deprecated/restricted permissions had minimal impact on performance and even improved accuracy in some models (e.g., CNN). Two strategies assessed whether this exclusion mitigates concept drift. Results showed improved drift detection--especially in a year-to-year setup—since deprecated permissions persisted in updated apps. Additionally, dataset balancing improved model accuracy and enhanced drift detection via the KS test.


\vspace{-10mm}
\begin{IEEEbiography}[{\includegraphics[width=1in,height=1.25in,clip,keepaspectratio]{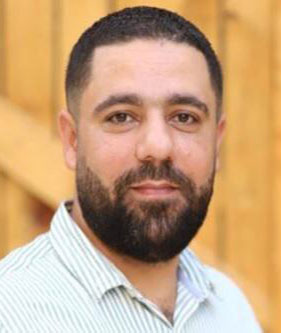}}]{Ahmed Sabbah} received a Bachelor's degree in computer science from An-Najah National University, Palestine, in 2008 and a Master's degree in software engineering from Birzeit University in 2021. He is currently working toward a Ph.D. degree with the Department of Computer Science, Birzeit University. His research interests include security, machine learning, software engineering, and mobile malware analysis.
\end{IEEEbiography}

\vspace{-10mm}
\begin{IEEEbiography}[{\includegraphics[width=1in,height=1.25in,clip,keepaspectratio]{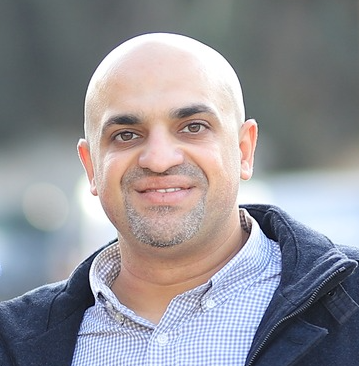}}]{Radi Jarrar} received his B.Sc. in Computer Information Technology from the Arab American University in 2007 and a Ph.D. in Computer Science from Monash University in 2012. Since 2015, he has been an assistant professor in the Department of Computer Science at Birzeit University, Ramallah, Palestine. His research interests include machine learning, computer vision, and data science, with applications in computer security.
\end{IEEEbiography}

\vspace{-10mm}
\begin{IEEEbiography}[{\includegraphics[width=1in,height=1.25in,clip,keepaspectratio]{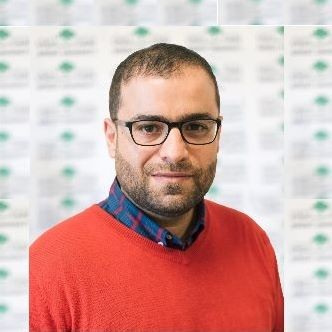}}]{Samer Zein} received the M.Sc. degree in Software Engineering from Northumbria University, United Kingdom, in 2004, and the Ph.D. degree in Mobile Software Engineering from the International Islamic University Malaysia (IIUM) in 2016. He is currently an Associate Professor in the Department of Computer Science at Birzeit University. He has over 20 years of academic experience and has contributed as a software engineer to several management information system (MIS) projects since 2000. His research interests include mobile app software engineering, empirical software engineering, model-driven software development, and systematic literature reviews (SLRs). He has also conducted multiple qualitative studies involving contemporary industrial case studies.
\end{IEEEbiography}

\vspace{-10mm}
\begin{IEEEbiography}[{\includegraphics[width=1in,height=1.25in,clip,keepaspectratio]{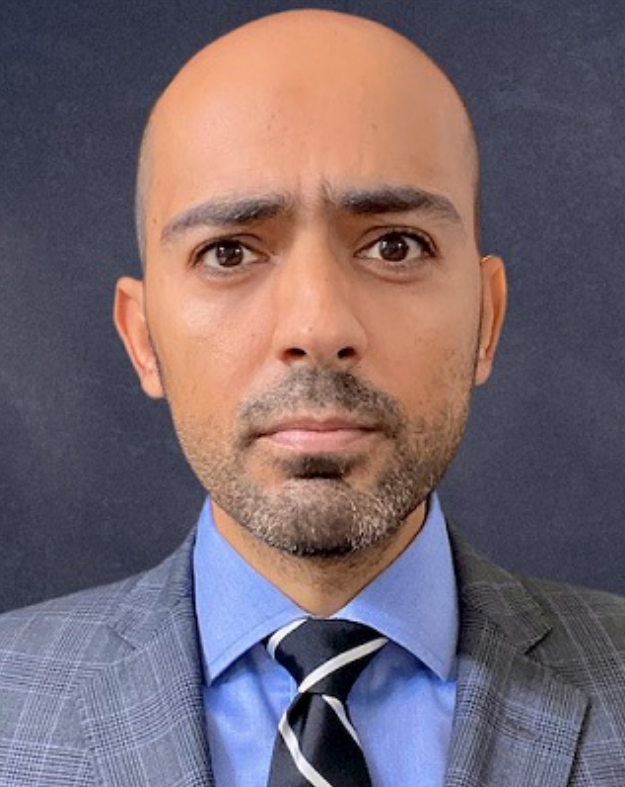}}]{David Mohaisen}
(Senior Member, IEEE) received the MSc and PhD degrees from the University of Minnesota in 2012. He is currently a full professor at the University of Central Florida, where he directs the Security and Analytics Lab. From 2015 to 2017, he was an assistant professor at SUNY Buffalo, and from 2012 to 2015, he was a senior research scientist with Verisign Labs. His research interests span networked systems security, online privacy, and measurements. He has been an associate editor for the IEEE Transactions on Mobile Computing, IEEE Transactions on Cloud Computing, IEEE Transactions on Parallel and Distributed Systems, and IEEE Transactions on Dependable and Secure Computing. He is a senior member of ACM (2018) and IEEE (2015), a distinguished speaker of the ACM, and a distinguished visitor of the IEEE Computer Society.
\end{IEEEbiography}

\end{document}